\documentclass[twocolumn,showpacs,preprintnumbers,amsmath,amssymb,prb]{revtex4-1}
\usepackage{graphicx}
\usepackage{dcolumn}
\usepackage{bm}
\usepackage{color}
\usepackage{ulem}
\usepackage{physics}

\renewcommand{\d}{{\bm{\delta}}}
\newcommand{\e}{{\bm e}}

\renewcommand{\k}{{\bm{k}}}
\newcommand{\q}{{\bm{q}}}
\renewcommand{\r}{{\bm{r}}}

\newcommand{\ii}{{\rm i}}

\def\lsim{\lower.35em\hbox{$\stackrel{\textstyle<}{\textstyle\sim}$}}
\def\gsim{\lower.35em\hbox{$\stackrel{\textstyle>}{\textstyle\sim}$}}

\begin{document}

\title{Kohn-Luttinger Superconductivity in Twisted Bilayer Graphene}

\author{J. Gonz\'alez$^{1}$ and T. Stauber$^{2}$}

\affiliation{
$^{1}$ Instituto de Estructura de la Materia, CSIC, E-28006 Madrid, Spain\\
$^{2}$ Materials Science Factory,
Instituto de Ciencia de Materiales de Madrid, CSIC, E-28049 Madrid, Spain}
\date{\today}

\begin{abstract}
We show that the recently observed superconductivity in twisted bilayer graphene (TBG) can be explained as a consequence of the Kohn-Luttinger (KL) instability which leads to an effective attraction between electrons with originally repulsive interaction. Usually, the KL instability takes place at extremely low energy scales, but in TBG, a doubling and subsequent strong coupling of the van Hove singularities (vHS) in the electronic spectrum occurs as the magic angle is approached, leading to extended saddle points in the highest valence band (VB) with almost perfect nesting between states belonging to different valleys. The highly anisotropic screening induces an effective attraction in a $p$-wave channel with odd parity under the exchange of the two disjoined patches of the Fermi line. We also predict the appearance of a spin-density wave (SDW) instability, adjacent to the superconducting phase, and the opening of a gap in the electronic spectrum from the condensation of spins
with wave vector corresponding to the nesting vector close to the vHS. 
\end{abstract}
 
\maketitle

{\it Introduction.} The discovery of superconductivity\cite{Cao18b} in twisted bilayer graphene (TBG) with a critical temperature of $1.7$ K at small twist angles around $1.1^\circ$ and Moir\'e-period of $\sim \!\!\! 13.5$ nm might be the missing puzzle needed to resolve long-standing questions related to high-$T_c$ superconductivity in layered compounds.\cite{Micnas90,Lee06,Stewart11} This hope is based on the fact that the phase diagram of TBG is characterised by a Mott-insulator at half-filling of the highest valence band (VB), corresponding to two electrons per Moir\'e unit cell, which upon doping turns into a superconducting (SC) instability.\cite{Cao18a} Increasing structural instead of chemical complexity can thus provide an alternative route to design devices with novel functionalities and therefore, TBG has attracted considerable interest even before the publication of Refs. \cite{Cao18a,Cao18b} due to its novel electronic\cite{Shallcross08,Schmidt10,Li10,Trambly10,Lee11,Luican11,Moon12,Sanchez-Yamagishi12,Brihuega12,Kim17,Zhang18}, optical,\cite{Moon13,Havener14,Patel15,Kim16,Suarez17,Stauber18} and plasmonic\cite{Stauber13,Stauber16,Hu17,Sunku18} properties. 


Although the findings by Jarillo-Herrero and co-workers have attracted immense attention,\cite{Xu18,Volovik18,Yuan18,Po18,Roy18,Guo18,Dodaro18,Baskaran18,Liu18,Huang18,Slagle18,Peltonen18,Kennes18,Koshino18,Kang18,Isobe18,You18,Wu18b,Zhang18,Pizarro18,Pal18,Ochi18,Thomson18,Carr18,Guinea18,Zou18} only very few attempts have focused on identifying the driving force of the superconductivity at the so-called magic twist angle $\theta_m\approx 1.05^\circ$, where the highest VB becomes extremely flat.\cite{Suarez10,Bistritzer11,SanJose12} One predictable theory was discussed in Ref. \onlinecite{Wu18b}, which sets the electron-phonon interaction as the basis of the pairing mechanism. On the other hand, the strong correlations that develop near the magic angle leave also room for the less conventional possibility of a purely electronic mechanism of superconductivity, following a route which has been also explored in the context of monolayer graphene.\cite{Uchoa07,BlackSchaffer07,Gonzalez08,Honerkamp08,Roy10,McChesney10,Nandkishore11,Guinea12,Gonzalez13} In this respect, there have been a couple of proposals in Refs. \onlinecite{Isobe18,You18} focusing on that kind of approach from a microscopic standpoint, investigating the weak-coupling instabilities arising from the shape of the Fermi surface (although for twisted bilayers relatively far away from the magic angle).

\begin{figure}
\includegraphics[width=0.49\columnwidth]{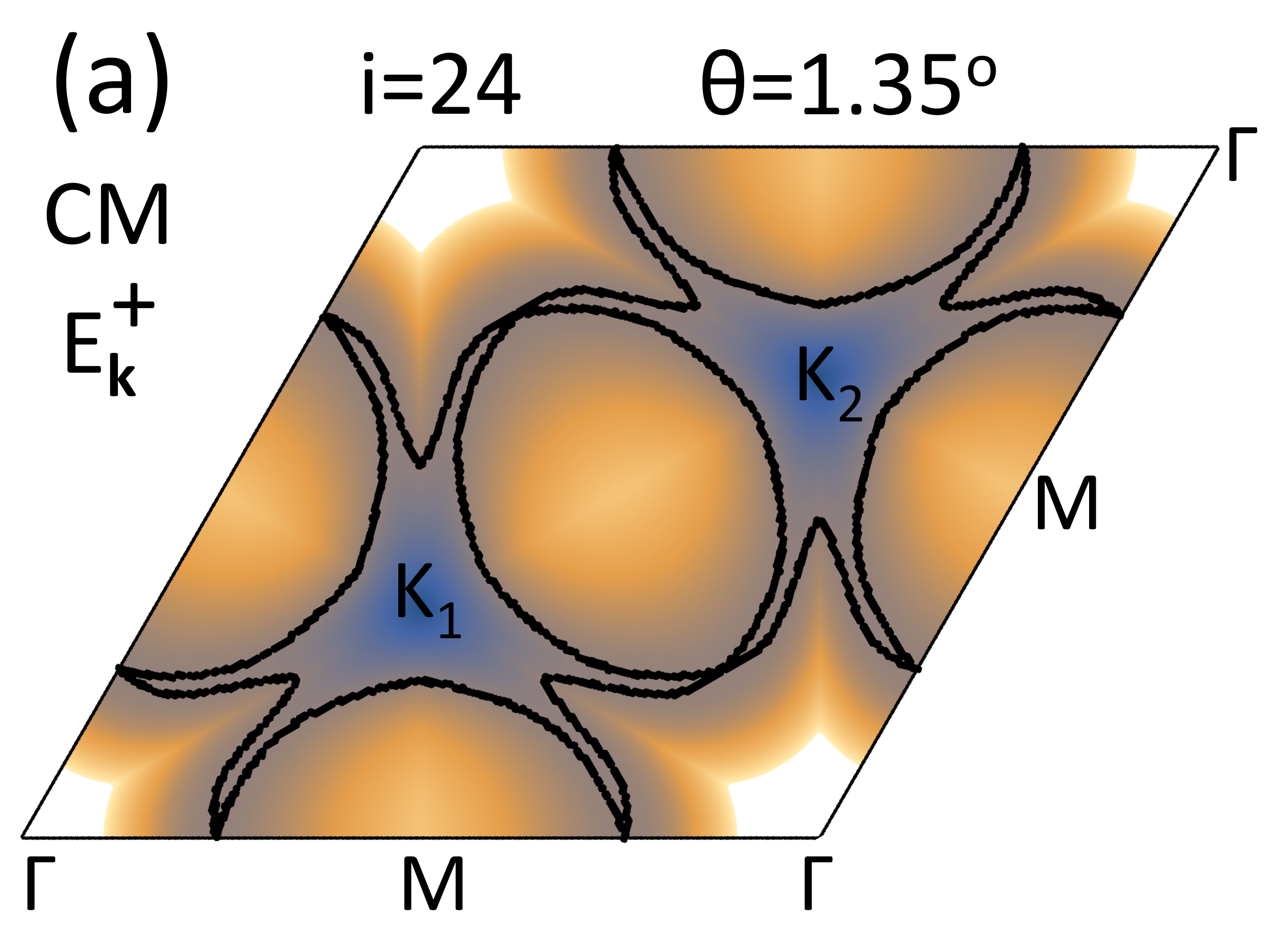}
\includegraphics[width=0.49\columnwidth]{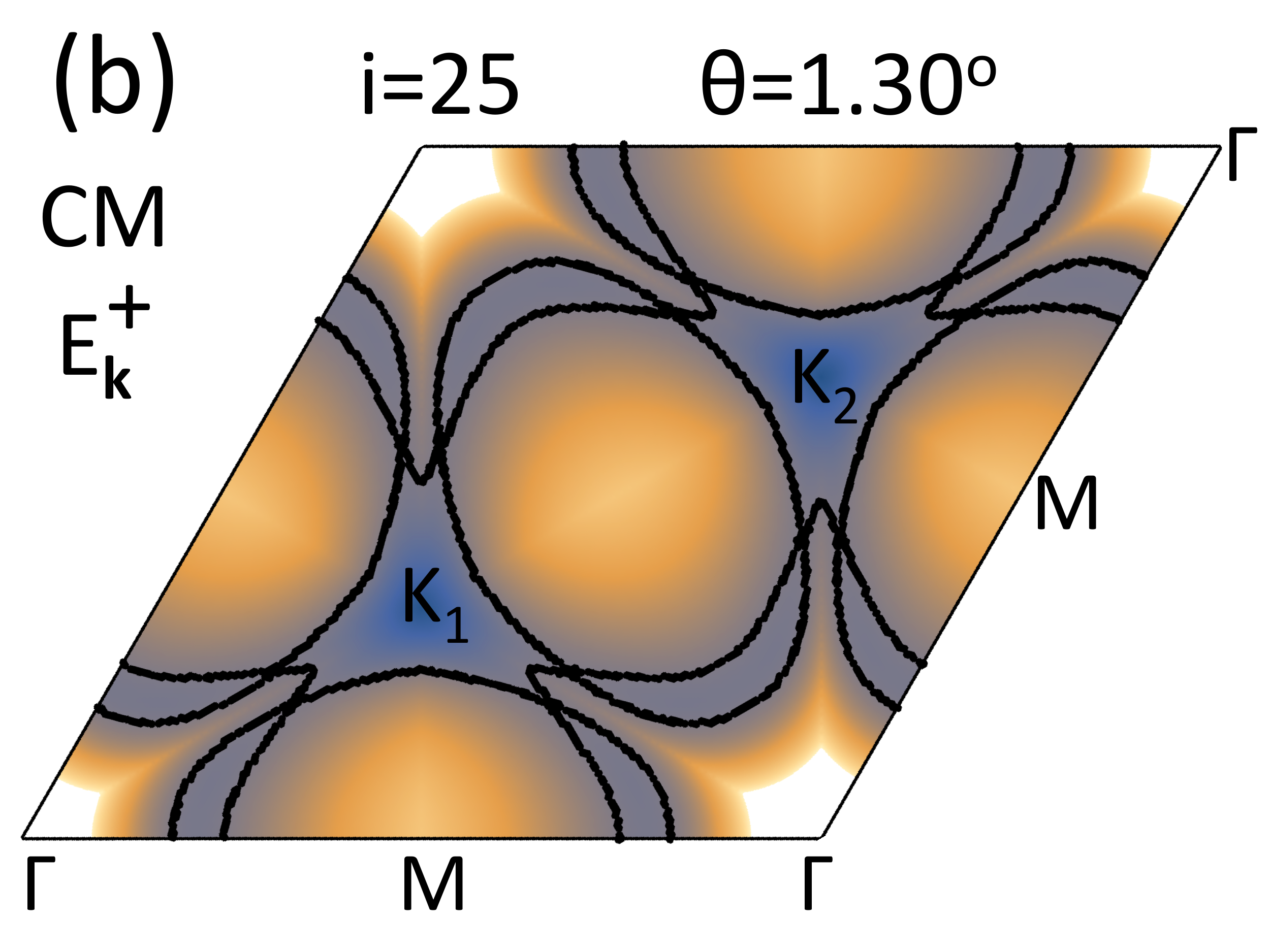}
\includegraphics[width=0.49\columnwidth]{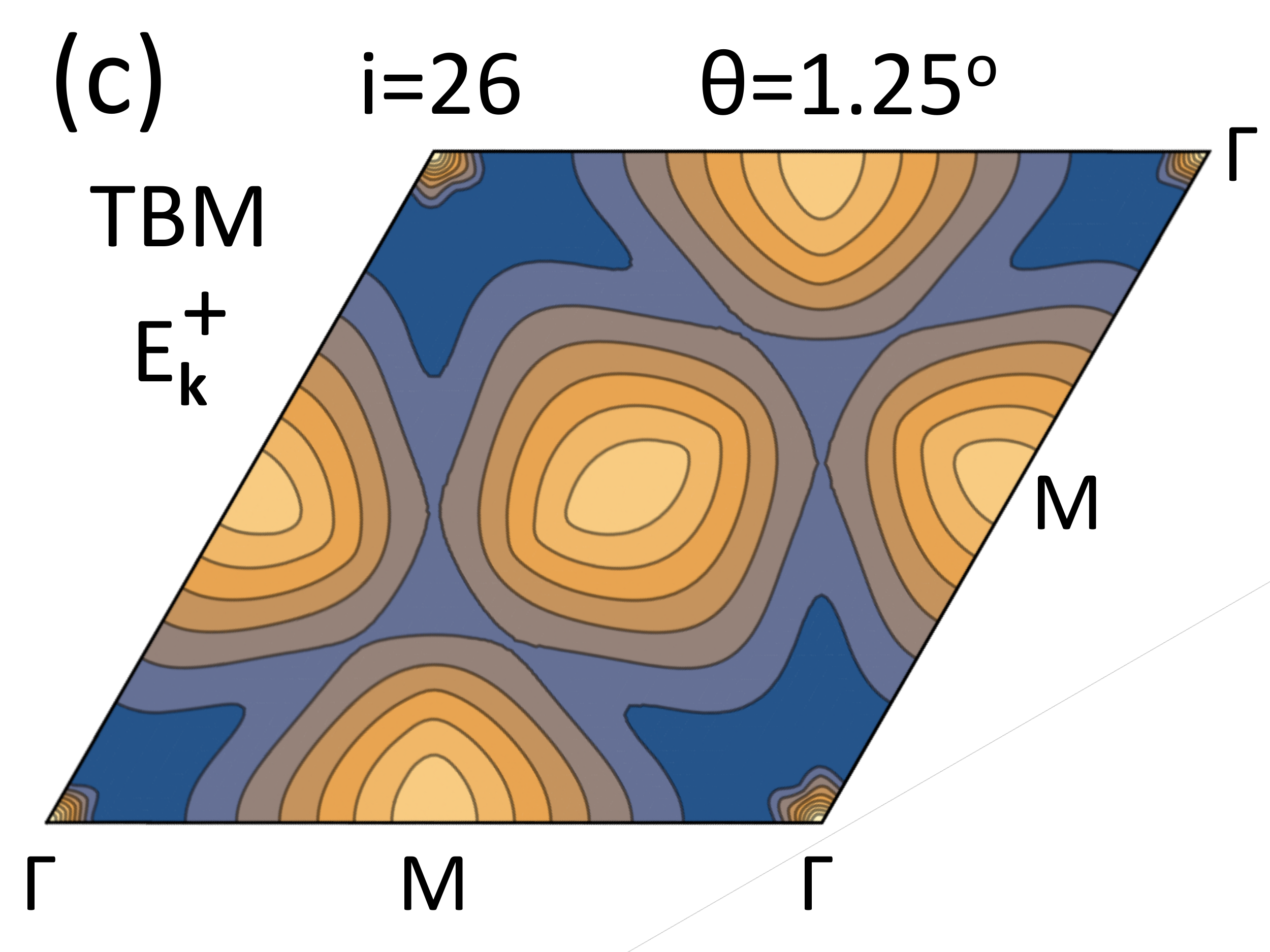}
\includegraphics[width=0.49\columnwidth]{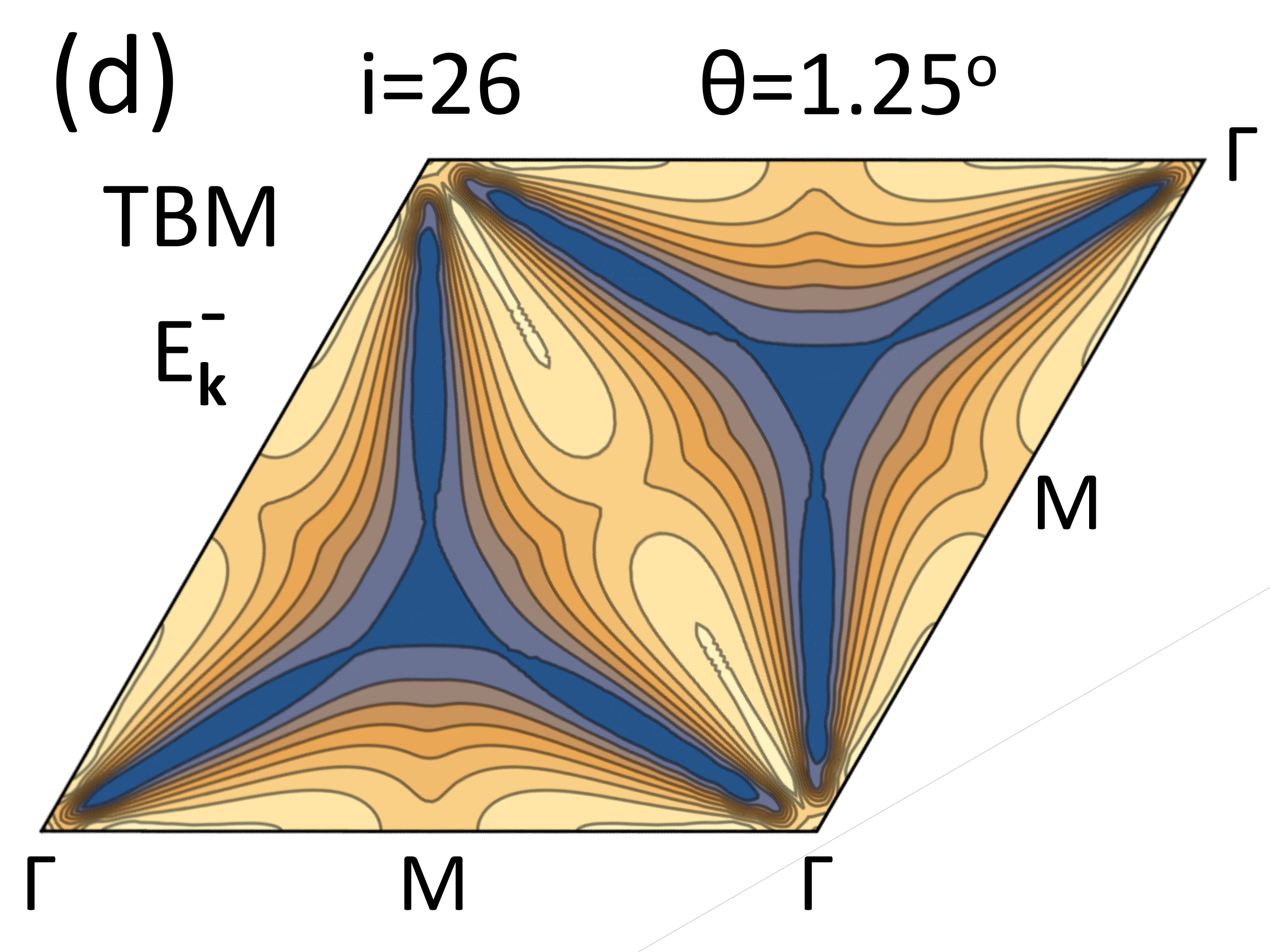}
\caption{(a) and (b): Density plot of the energy dispersion of the highest valence band $E_\k^+={\rm{max}}(E_\k^K,E_\k^{K'})$ in the Moir\'e Brillouin zone (MBZ) of the continuous model for two different twist angles. Dark (bright) colors represent high (low) energies and the black contour lines represent the Fermi surface at the energy of the van Hove singularity (vHS), $E_{\rm{vH}}$. There occurs a doubling of the vHS at some critical angle $\theta_{i=24} > \theta_c^+ > \theta_{i=25}$, i.e., for lower $\theta $ there are twelve saddle points located inside the MBZ close to the lines that connect the $\Gamma$ and  $K_\ell$-points. (c) and (d): Contour plot of the highest valence bands $E_\k^+$ and $E_\k^-$ of the tight-binding model for  ${i=26}$. The two sets of vHS belonging to different valleys have already merged and are found now in different bands $E_\k^+$ and $E_\k^-$. \label{one}}
\end{figure}

In this Letter, we unveil what may be the key interaction governing the superconductivity of TBG, identifying for that purpose a number of universal topological features in the electronic dispersion of the the highest VB which are indispensable to understand the pairing mechanism at the microscopic level. We show that, below a certain critical twist angle $\theta^-_c \approx 1.3^\circ$, there is both a doubling and strong coupling of the van Hove singularities (vHS) in the electronic spectrum, leading to extended saddle points in the highest VB with almost perfect nesting between states belonging to different valleys. This induces a highly anisotropic screening of the Coulomb interaction, leading necessarily to an effective attractive interaction in a channel with $p$-wave symmetry which is the seed required to trigger the superconducting instability.

Our theoretical construction constitutes a variant of the so-called Kohn-Luttinger (KL) mechanism\cite{Kohn65,Baranov92} which was proposed as a route to develop a superconducting instability starting from a purely repulsive interaction. We thus put forward a microscopic theory of superconductivity in TBG which only relies on the Coulomb interaction, giving definite quantitative predictions for the critical energy scale of the superconducting transition in a range which spans from weak-coupling up to a much stronger instability depending on the proximity of the Fermi level to the vHS. We also complete the study by discussing the spin-density wave (SDW) adjacent to the superconducting phase, and whose onset takes place typically for a critical Coulomb interaction which is below the bandwidth of the highest VB, thus reassuring our microscopic approach to the superconductivity of TBG.

{\it Models.} To model TBG, we will use the continuous model (CM) that treats commensurate lattices parametrised by the integer $i$ with the twist angle $\cos\theta_i=\frac{3i^2+3i+0.5}{3i^2+3i+1}$.\cite{Lopes07,Mele10,Bistritzer11,Moon12} For these angles, we will also use the tight-binding model (TBM) of TBG\cite{Suarez10,Trambly10,Moon13} which has already built in the coupling between states around the $K$-valley and their time-reversed partners in the $K'$-valley. In the Supplementary Material (SM),\cite{SI} the real space image and the Brillouin zones of the two layers are shown together with the Moir\'e Brillouin zone (MBZ) around the two valleys $K$ and $K'$.

In the TBM, the highest VB containing up to four electrons (corresponding to twofold spin- and valley-degeneracy as mentioned in the introduction) thus splits in two bands. Consequently, the TBM description can be compared to the CM by combining in the latter the highest VB corresponding to each $K$-point, $E_\k^K$ and $E_\k^{K'}$, to $E_\k^+={\rm{max}}(E_\k^K,E_\k^{K'})$ and $E_\k^-={\rm{min}}(E_\k^K,E_\k^{K'})$. The result of the comparison turns out to be in general quite satisfactory, as shown in the SM. For twist angles $\theta>1.1^\circ$, the two combined bands $E_\k^+$ and $E_\k^-$ are only degenerate on the six $\Gamma K_\ell$-lines for which $E_\k^K=E_\k^{K'}$ in the CM, $K_\ell$ being the Dirac point belonging to layer $\ell=1, 2$ and $\k$ measured with respect to the corresponding valley. Also the density plot of the highest VB and degeneracy contours for smaller twist angles $\theta<1.1^\circ$ are discussed in the SM.

{\it Van Hove singularities in the highest VBs.}
The KL mechanism we are proposing relies on the anisotropic screening that can be provided by a strong vHS, induced by a large number of saddle points within the  MBZ  which is crucial to tip the scale towards a SC instability.\cite{Nandkishore11}  In this regard, the highest VB $E_\k^+$ witnesses important changes with respect to its topology as function of the twist angle. At large twist angles, there are six vHS (saddle points), three for each valley and located around the three $M$-points of the MBZ. Decreasing the twist angle, the vHS move away from the $M$-points and for $i\approx 24-25$, we observe a splitting of the saddle points, see Fig. \ref{one} (a) and (b) where the density plot of $E_\k^+$ is shown together with the Fermi line at the vHS. We thus identify a first critical angle $\theta_c^+$, where a doubling of vHS occurs from six to twelve.  The exact crossing point usually occurs at a non-commensurate critical angle $\theta_c^+$ that can be treated by more advanced numerical techniques.\cite{Carr17,Cances17} 

At smaller twist angles $\theta<\theta_c^+$, the evolution of the saddle points critically depends on the coupling between states at different $K$ and $K'$ valleys, best captured by the TBM. Decreasing the twist angle, the pairs of vHS move closer to the $\Gamma K_\ell$-lines up to a second critical angle $\theta_c^-$ at which the two saddle points of the $E_\k^+$-band merge and and a new saddle point in the $E_\k^-$-band emerges. Both vHS are then pinned to the $\Gamma K_\ell$-line, see Fig. \ref{one} (c) and (d) where the contour plot of $E_\k^+$ and $E_\k^-$ is shown for $i=26$. 

The splitting of the six pairs of vHS has two important consequences. First, the overlap between the states around the saddle points becomes approximately one due to the direct coupling of the two valleys; second, the vHS become further extended. Both consequences lead to a large susceptibility of particle-hole pairs when the Fermi energy is close to the vHS, which can trigger the SC instability via the KL mechanism. 

\begin{figure}
\includegraphics[width=0.99\columnwidth]{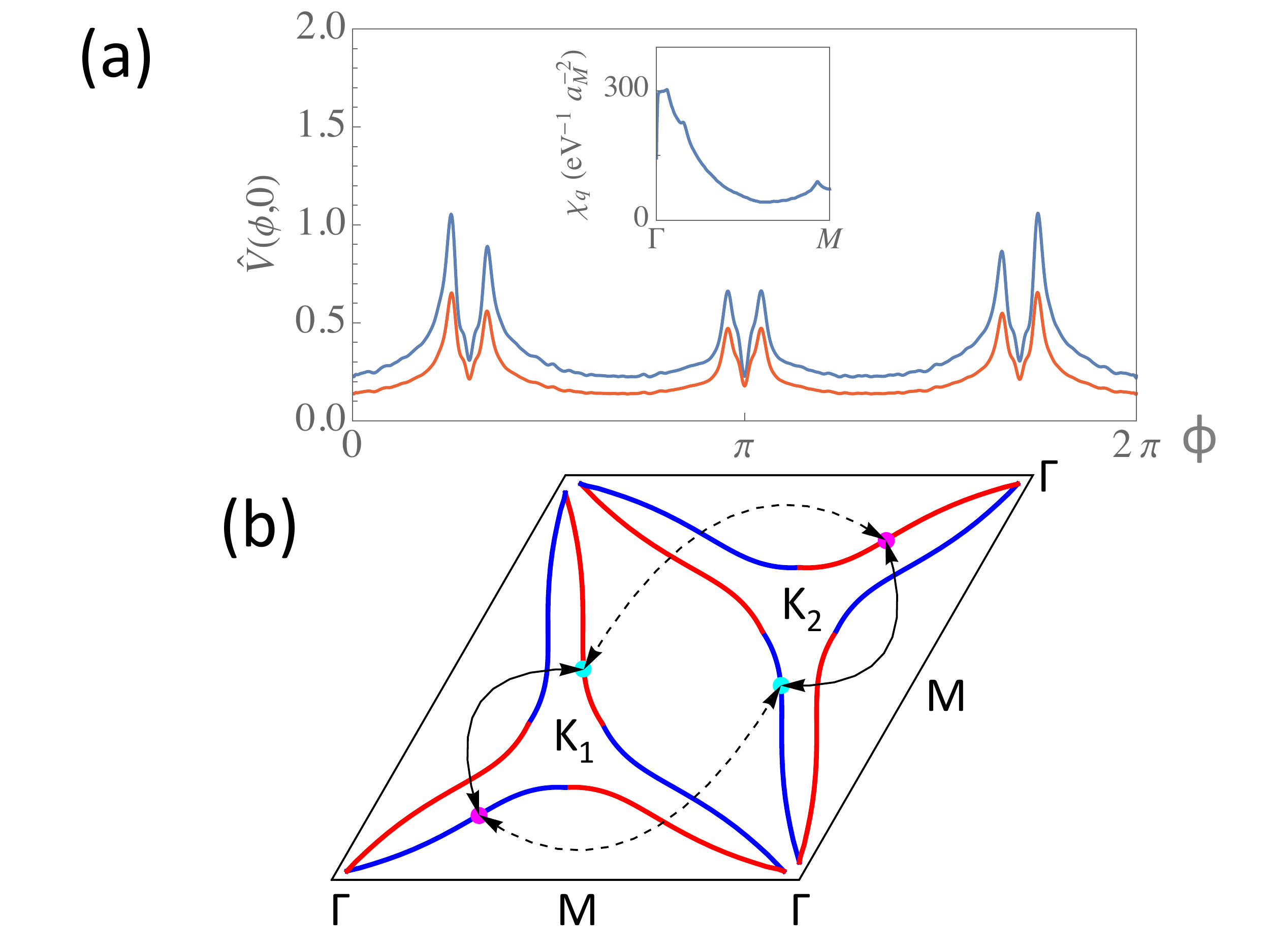}
\caption{(a) Plot of the BCS vertices $\widehat{V}_{\rm inter} (\phi, 0)$ (blue 
curve) and  $\widehat{V}_{\rm intra} (\phi, 0)$ (red curve) for angle $\phi $ running along one of the patches of the Fermi line of a twisted bilayer with $i = 26$ and Fermi level placed 0.1 meV below the vHS of the valence band $E_\k^-$, for a value of the Coulomb interaction $U/a_M^2 = 2$ meV ($a_M$ being the Moir\'e lattice constant of the twisted bilayer graphene). The inset shows the particle-hole susceptibility $\chi_\q$ (in units of eV$^{-1}a_M^{-2}$) for momenta running from $\Gamma $ to M in the MBZ with the same parameters. (b) Intrapatch (solid lines) and interpatch (dashed lines) scattering processes of two Cooper pairs (cyan and magenta dots). The blue respectively red curves indicate the states belonging to the two different valleys of the Fermi line $0.2$ meV below the energy of the vHS of $E_\k^-$ as obtained from the CM with $i=29$. }
\label{two}
\end{figure}

{\it Kohn-Luttinger instability.} The KL instability can be analyzed starting from a conventional BCS approach 
where the Cooper-pair vertex $V$ is parametrized in terms of the 
angles $\phi $ and $\phi' $ of the respective momenta of the spin-up 
incoming and outgoing electrons on each contour line of energy $\varepsilon$.
The iteration of the scattering between the electrons in the Cooper pair can
be encoded in the self-consistent equation
\begin{eqnarray}
&&V(\phi, \phi')= V_0 (\phi, \phi')-   \nonumber        \\ 
 && \frac{1}{(2\pi )^2} \int^{\Lambda_0} \frac{d \varepsilon }{\varepsilon } 
   \int_0^{2\pi } d \phi'' 
  \frac{\partial k_\perp }{\partial \varepsilon}  
      \frac{\partial k_\parallel }{\partial \phi''} 
  V_0 (\phi, \phi'')    
         V(\phi'', \phi')
\label{pp}
\end{eqnarray}
where $k_\parallel ,k_\perp $ are the respective longitudinal and transverse 
components of the momentum while $V_0 (\phi, \phi') $ stands for the bare
vertex at a high-energy cutoff $\Lambda_0$. Differentiating Eq. (\ref{pp}) 
with respect to the cutoff, we end up with the scaling equation
\begin{equation}
\Lambda \frac{\partial \widehat{V}(\phi, \phi' )}{\partial \Lambda } 
 =   \frac{1}{2\pi }  \int_0^{2\pi } d \phi''  
 \widehat{V} (\phi, \phi'' )  \widehat{V}(\phi'', \phi' )
\label{scaling}
\end{equation}
where
$\widehat{V} (\phi, \phi' ) = F(\phi ) F(\phi' ) V (\phi, \phi' )$
and $F(\phi ) = \sqrt{ (\partial k_\perp / \partial \varepsilon )
  (\partial k_\parallel / \partial \phi )/2\pi  }$.
It is clear that, if $\widehat{V}(\phi, \phi' )$ has some negative 
eigenvalue at the high-energy regime of $\Lambda $, this will result in a 
divergent growth of the BCS vertex in the low-energy limit 
$\Lambda \rightarrow 0$, which is the signature of a pairing instability.

The KL mechanism of superconductivity is enhanced for electron systems in which the Fermi velocity has a large anisotropy along the Fermi line. The anisotropic screening induced by particle-hole excitations gives rise to the angular dependence of the BCS vertex which, assuming a constant interaction $U$ in momentum space,\cite{SI} becomes in the random-phase approximation (RPA)\cite{Scalapino87}
\begin{equation}
V_0 (\phi, \phi') =  U +
  \frac{U^2 \chi_{\k+\k^\prime}}
         {1 - U \chi_{\k+\k^\prime}}
 + \frac{U^3 \chi_{\k-\k^\prime}^2}
       {1 - U^2 \chi_{\k-\k^\prime}^2} \;,
\label{init}
\end{equation} 
where $\k, \k^\prime$ are the respective momenta at angles 
$\phi, \phi'$ and $\chi_{\q}$ is the particle-hole susceptibility at momentum 
transfer $\q$.

A simple argument allows us to understand why there is always an effective 
attractive interaction in TBG approaching the magic angle: since the Fermi line near the vHS in the VB $E_\k^-$ consists of two disjoined patches, the two electrons forming the Cooper-pair belong to different patches, see Fig. \ref{two} (b). We can now distinguish between two different contributions to the BCS vertex, depending on whether the electrons of the Cooper pair scatter within the same patch of the Fermi line (intrapatch vertex $V_{\rm intra}$)
or whether they scatter exchanging their patches (interpatch vertex $V_{\rm inter}$).
Close to the vHS of the VB $E_\k^-$, the particle-hole susceptibility has a large peak at small momentum transfer (as seen in the inset of Fig. \ref{two} (a)) which leads to a strong enhancement of the second term on the right hand side of Eq. (\ref{init}) for $V_{\rm inter}$, when 
$\k \approx -\k^\prime$. This enhanced susceptibility can be understood from the almost perfect nesting condition that connects the two opposite lines of the three side lobes of each patch, as depicted in Fig. \ref{two} (b). In the case of $V_{\rm intra}$, however, the enhancement corresponds to the third term in the equation when $\k \approx \k^\prime$, which is of lower order than the large contribution picked by $V_{\rm inter}$, as shown in Fig. \ref{two} (a).

The full BCS vertex $\widehat{V} (\phi, \phi' )$ now
becomes a matrix such that 
\begin{equation}
\widehat{V} = \left(\begin{array}{cc}   \widehat{V}_{\rm intra}   &   \widehat{V}_{\rm inter}     \\  
        \widehat{V}_{\rm inter}   &   \widehat{V}_{\rm intra}          \end{array} \right)\;.
\end{equation}
Given that the interpatch scattering is in general more intense than the  
intrapatch interaction, $\widehat{V}_{\rm inter} \gtrsim \widehat{V}_{\rm intra}$, 
we find an attractive channel with negative eigenvalue and antisymmetric amplitude in the two disjoined patches of the Fermi line, see also SM.\cite{SI}  
\begin{figure}
\includegraphics[width=0.99\columnwidth]{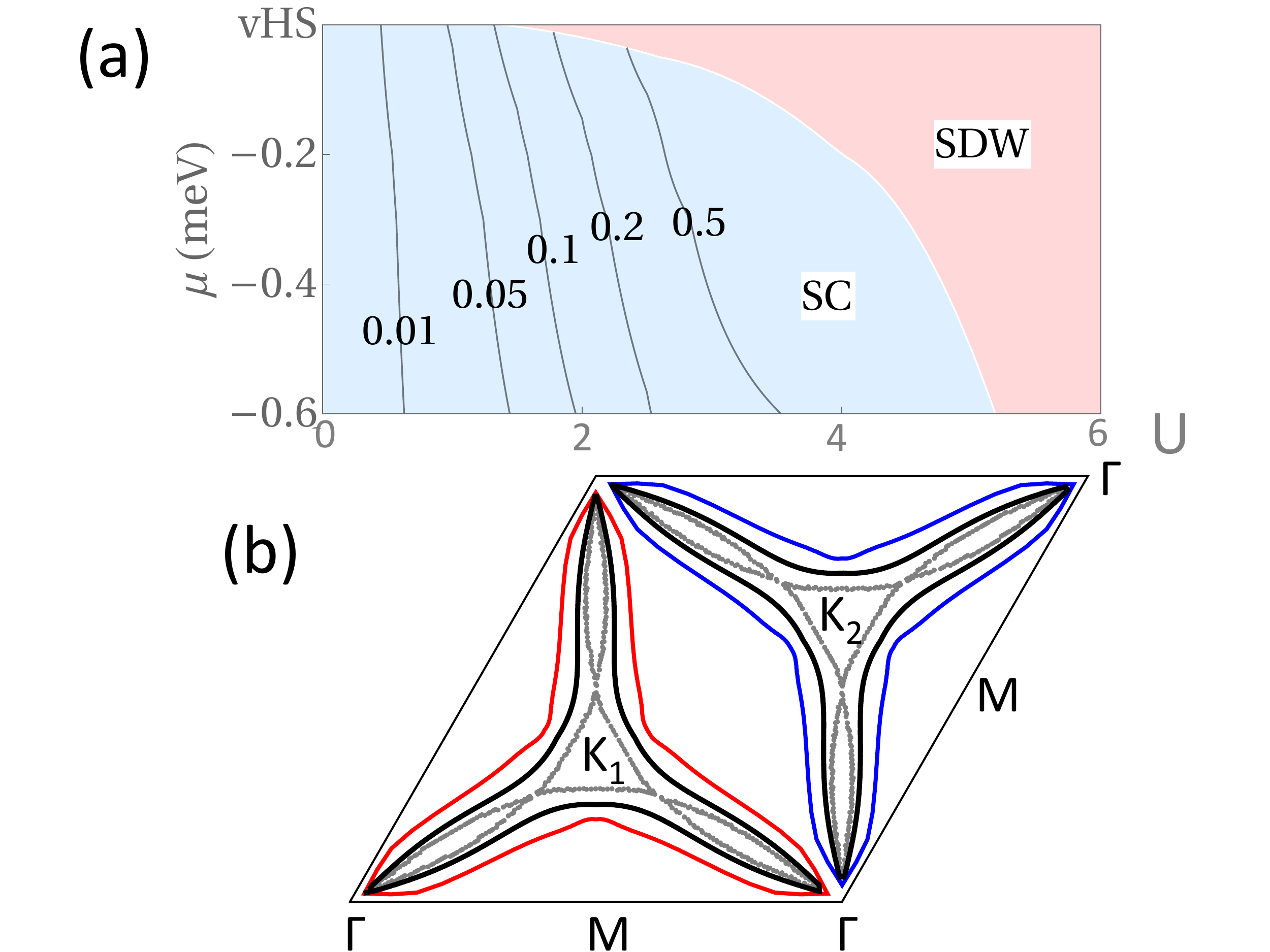}
\caption{(a) Phase diagram as function of the chemical potential $\mu $ relative to the energy at the vHS of $E_\k^-$ and the bare interaction $U$ (in units of meV $a_M^2$) . The superconducting instability (SC) always precedes the spin-density-wave instability (SDW). Contour lines refer to the largest value of the BCS coupling $|\lambda|$ in the channel with dominant attractive interaction. (b) Gap structure with $\Delta_\phi=\Delta_0(0.75-0.25\cos(3\phi))$ as red and blue curves to indicate the sign change under parity. The gap of $\Delta_0=8$ K is exaggerated in order to demonstrate the qualitative behavior, i.e., the gap region is larger around regions where screening is enhanced. Also shown the Fermi line at the vHS (grey dots) and the Fermi line $0.2$ meV below the energy of the vHS (black).}
\label{three}
\end{figure}

{\it Broken symmetry phases.} The poles in the RPA expression in Eq. (\ref{init}) imply the existence of a 
critical interaction strength at which the BCS vertex as well as other response 
functions diverge, indicating the trend towards broken symmetry phases in the electronic system. 
The competition between these low-energy phases can be analyzed in an unbiased manner by means of a renormalization group (RG) approach, see SM.\cite{SI} It turns out that there is a phase boundary between the pairing instability and a spin-density wave instability which prevails above a critical interaction $U_c$, as shown in Fig. \ref{three} (a). The value of $U_c$ is dictated by the peak of the particle-hole susceptibility, $\chi_\q$, whose position in momentum space sets the wave vector of the spin-density wave instability - as seen in the inset of Fig. \ref{two} (a). We expect that in the strong coupling regime, the spin-density wave instability, which also opens a gap in the electronic spectrum,\cite{Gonzalez02,Gonzalez03} should correspond to the insulating phase observed in the experiments of Ref. \onlinecite{Cao18b}.
Nevertheless, we stress that the KL instability is always dominant before reaching the critical interaction $U_c$, as it only relies on the anisotropy of the BCS vertex.

In our RG approach, $U$ corresponds to the interaction potential at zero momentum, i.e., the scale-invariant part of the interaction close to the vHS, see SM.\cite{SI} As shown in Fig. \ref{three} (a), the values of $U_c$ are of the order of a few meV (times the square of the Moir\'e lattice constant $a_{M}$). Those values match well with the order of magnitude expected for the Coulomb repulsion, which must undergo a reduction (with respect to that in monolayer graphene) by a factor inversely proportional to the number of atoms in the unit cell of the twisted bilayer. 
The values of $U/a_{M}^2$ needed to trigger the broken symmetry phase (and thus a preceding KL instability) are therefore 
below the bandwidth  $W\sim 5$ meV of the highest VB of the twisted bilayers considered here ($i=26$). 

\begin{table}[t]
\begin{tabular}{c|c|c}
Eigenvalue $\lambda$ &Irreducible Representation&Parity\\
\hline
-0.51&$A_1$&Odd\\
-0.11&$E$&Odd\\
-0.10&$A_2$&Even\\
-0.08&$A_1$&Odd\\
-0.06&$E$&Even
\end{tabular}
\caption{Most negative (attractive) eigenvalues and their respective irreducible representations of $C_{3v}$ for $U/a_M^2=2.5$ meV and the Fermi level placed $0.1$ meV below the vHS in the valence band $E_\k^-$.} 
  \label{table}
\end{table}

{\it Superconducting order parameter.} 
Fourier transforming the BCS vertex, we can identify the attractive channels with negative eigenvalues $\lambda$ and their respective symmetries. We recall that, for each disjoined patch of the Fermi line with $K_\ell$ as its centre point, the relevant point group is $C_{3v}$. The irreducible representations of $C_{3v}$ can be characterised by the Fourier components, i.e., $A_1\rightarrow\{\cos (3n\phi)\}$, $A_2\rightarrow\{\sin (3n\phi)\}$, and $E\rightarrow\{\cos (m\phi),\sin (m\phi)\}$ with $m\neq3n$, $\phi=0$ corresponding to the point on the Fermi line closer to $K_\ell$. 

Interestingly, we find attractive channels belonging to all three irreducible representations with odd {\em and} even parity, see Table \ref{table}. But the dominant instability is given by an order parameter that transforms according to $A_1$ with odd parity and can be approximated by $\Delta_\phi=\Delta_0[0.75-0.25\cos(\phi)]$. The values of the most attractive coupling $|\lambda|$ are represented in Fig. \ref{three} (a).
The scale of the gap $\Delta_0$ is obtained by solving Eq. (\ref{scaling}) and yields
\begin{equation}
\Delta_0=\Lambda_0\exp\left(-1/|\lambda| \right)\;.
\end{equation}
We can approximate the cutoff scale $\Lambda_0$ by the separation of the Fermi line from the energy of the vHS, i.e., $\Lambda_0\sim0.1$ meV which is of the order of 1 K.
The gap structure is shown in Fig. \ref{three} (b) for $\Delta_0=8$ K. The order parameter is slightly suppressed close to the $K_\ell$-points and changes sign under parity indicated by the blue and red lines, respectively. Also shown is the Fermi line $0.2$ meV below the $E_{\rm vH}$ (black) as well as the Fermi line at the vHS containing the six saddle points (gray dots). 

{\it Summary.}
We present a quantitative theory for the recently discovered superconductivity in twisted bilayer graphene close to, {\it but not at} the first magic angle. Our theory rests upon the observation that there is a saddle-point splitting at some critical twist angle which induces strong intervalley coupling. Fermi lines close to the vHS are disjunct and display regions of almost perfect nesting giving rise to a large susceptibility for small wave numbers and thus to an enhanced Kohn-Luttinger instability. The dominant instability yields an order parameter with odd parity that has an approximate $s$-wave symmetry around the two patches of the Fermi line, i.e., it is spin-triplet and valley-singlet. The sign-change of the superconducting gap for different valleys should be detectable via STM by measuring the quasiparticle interferences.
Furthermore, our theory predicts a scale of the superconducting gap which agrees with the experimental findings of Ref. \onlinecite{Cao18b}. 

Let us finally address open questions. In order to find the superconducting instability in the middle of the VB, the level of the vHS should correspond to half-filling. This is indeed the case for angles in the vicinity (but not at) the magic angle and there is further experimental evidence of an interaction induced pinning of the vHS to half-filling,\cite{Kim16b,Cao16} see SM.\cite{SI} For twist angles in the immediate vicinity of the magic angle, our computational scheme breaks down there, but we believe that the key features we have found driving the KL mechanism must also be present in that regime, possibly yielding an even larger superconducting gap. 

{\it Acknowledgements.}
We thank G. G\'omez-Santos and H. Kohler for useful discussions. This work has been supported by Spain's MINECO under Grant No. FIS2017-82260-P.

\begin{widetext}
{\bf\Huge SUPPLEMENTARY INFORMATION}
\section{Continuous Hamiltonian}
\label{ContinuousHamiltonian}

We use the following continuous Hamiltonian to analyze the band structure of twisted bilayer graphene (TBG)\cite{Lopes07,Bistritzer11}
\begin{align}\label{Hamiltonian}
\mathcal{H } =&\hbar v_F\sum_{\bm k} c_{1,\bm k,\alpha}^{\dagger} \;\bm \tau_{\alpha\beta}^{-\theta/2} \cdot (\bm k + \frac{\Delta\bm K}{2}) \; c_{1,\bm k,\beta} \notag\\ 
+&\hbar v_F\sum_{\bm k} c_{2,\bm k,\alpha}^{\dagger} \;\bm \tau_{\alpha\beta}^{+\theta/2} \cdot (\bm k - \frac{\Delta\bm K}{2}) \; c_{2,\bm k,\beta}\\
+&\frac{t_{\perp}}{3} \sum_{\bm k,\bm G} (c_{1,\bm k + \bm G,\alpha}^{\dagger} \; T_{\alpha\beta}(\bm G) \; c_{2,\bm k,\beta} + H. c.)\notag\;,
\end{align}
where $( \bm \tau^{\gamma}_x,\bm \tau^{\gamma}_y ) =  e^{\ii\gamma\bm\tau_z/2} ( \bm \tau_x, \bm \tau_y ) e^{-\ii\gamma\bm\tau_z/2}  $, $\bm \tau_{x,y,z}$ being Pauli matrices. The Dirac cones are separated by $\Delta \bm K = 2 |\bm K| \sin(\theta/2)  \left[ 0,1\right]$  with $\bm K = \tfrac{4 \pi}{3 a_g} \left[ 1,0\right] $. The interlayer hopping is restricted to wave vectors $\bm G = \{\bm 0,-\bm G_1,-\bm G_1-\bm G_2\} $ with $\bm G_1 =  |\Delta \bm K| \left[ \tfrac{\sqrt{3}}{2},\tfrac{3}{2}\right]$, $\bm G_2 =  |\Delta \bm K| \left[ -\sqrt{3},0\right]$, and
\begin{equation}\label{Hopping}
T(\bm 0) = \begin{bmatrix} 1 & 1\\1 & 1\end{bmatrix}; \; \; T(-\bm G_1) = T^{*}(-\bm G_1-\bm G_2)=\begin{bmatrix} 
e^{\ii 2 \pi / 3} & 1\\ e^{-\ii 2 \pi / 3}& e^{\ii 2 \pi / 3}\end{bmatrix}
.\end{equation}
The Hamiltonian for the other valley is related to Eq. (\ref{Hamiltonian}) via time reversal symmetry, see e.g. Ref. \cite{Moon13}. The real space image of TBG and the two Brillouin zones of the uncoupled graphene layers (red and blue hexagons)  together with the two Mo\'ire Brillouin zones (MBZs) of TBG (thick black rhombuses) are shown in Fig. \ref{TBL} (a) and (b), respectively.

For the calculations, we use the hopping amplitudes $t=-2.78\,\text{eV}$ and $t_{\perp}=0.33\,\text{eV}$ with $\hbar v_F= \tfrac{\sqrt{3}}{2} t a$ and $a=2.46 \, \mathring{\text{A}}$. Twist angles have been chosen from the set of commensurate structures labeled by $\cos(\theta_i)=1-\tfrac{1}{2(3i^2 + 3i + 1)}$. For the above parameters, the magic angle occurs at $i\approx31$.

\section{Tight-binding Hamiltonians}
For the accurate calculation of the effective Cooper-pair vertex, the use of the tight-binding model is indispensable. This approach accounts for the approximate nesting between electronic states belonging to different valleys, which can lead to a large susceptibility only if such states have nonvanishing overlap. We adopt a general formulation of the tight-binding approach with Hamiltonian:
\begin{align}
H = -\sum_{\langle i,j\rangle} t_{\parallel} (\r_i-\r_j) \; (a_{1,i}^{\dagger}a_{1,j}+h.c.) - \sum_{\langle i,j\rangle} t_{\parallel} (\r_i-\r_j) \; (a_{2,i}^{\dagger}a_{2,j}+h.c.) - \sum_{(i,j)} t_{\perp}(\r_i-\r_j) \; (a_{1,i}^{\dagger}a_{2,j}+h.c.)\;. 
\label{tbh}
\end{align}
The sum over the brackets $\langle...\rangle$ runs over pairs of atoms in the same layer (1 or 2), whereas the sum over the curved brackets $(...)$ runs over pairs with atoms beloging to different layers. $t_{\parallel} (\r)$ and $t_{\perp} (\r)$ are hopping matrix elements which have an exponential decay with the distance $|\r|$ between carbon atoms. A common parametrization is based on the Slater-Koster formula for the transfer integral\cite{Moon13} 
\begin{align}
-t(\d)=V_{pp\pi}(d)\left[1-\left(\frac{\d\cdot\e_z}{d}\right)^2\right]+V_{pp\sigma}(d)\left(\frac{\d\cdot\e_z}{d}\right)^2
\end{align}
with
\begin{align}
V_{pp\pi}(d)=V_{pp\pi}^0\exp\left(-\frac{d-a_0}{r_0}\right)\;,
V_{pp\sigma}(d)=V_{pp\sigma}^0\exp\left(-\frac{d-d_0}{r_0}\right)\;,
\end{align}
where $\d $ is the vector connecting the two sites, $\e_z$ is the unit vector in the $z$-direction, $a_0 $ is the C-C distance and $d_0$ is the distance between layers. A typical choice of parameters is given by $V_{pp\pi}^0=-2.7$ eV, $V_{pp\sigma}^0=0.48$ eV and $r_0=0.319 a_0$ \cite{Moon13}. In particular, we have taken these values to carry out the analysis shown below about the evolution of the saddle points in the highest valence bands of the twisted bilayers. For an alternative comparison between continuous and tight-binding model, see Ref. \cite{Stauber18b}.

On the other hand, a different point of view consists in thinking about the hopping parameters as phenomenological variables which have to be adjusted to reflect the details of the energy bands. In this respect, the hopping parameters show less regularity for large distance between carbon atoms, and they may even alternate sign when increasing their separation. We have probed the dependence on the tight-binding parametrization by adopting an alternative choice of hopping matrix elements, constraining $t_{\parallel} (\r)$ to $t_0 = 3$ eV for nearest-neighbor hopping and setting the decay for interlayer hopping   
\begin{align}
-t_{\perp } (\d) = t_{\perp 0 }  \exp\left(-\frac{d-d_0}{r_0}\right)
\end{align}
with $t_{\perp 0 } = 0.296$ eV and $r_0 = 0.076$ nm (and truncating in practice the decay for separation $d > 2.43 a_0$). Despite the rather different choice of parameters with respect to the values taken above, we have checked that this formulation of the tight-binding model leads to the same predictions described in Sec. III regarding the evolution of the highest valence bands in the twisted bilayers, including the topological transitions at critical angles $\theta_c^+$ and $\theta_c^-$. We have adopted this phenomenological choice of parameters (which accounts in particular for sensible values of both the nearest-neighbor hopping $t_0$ and interlayer hopping $t_{\perp 0 }$) to carry out the calculations reported in the main text, relying on a topology of the Fermi line shown there in Fig. 1(d) which is quite similar to that represented in Fig. 7 c) below.

\begin{figure*}
\includegraphics[width=0.3\columnwidth]{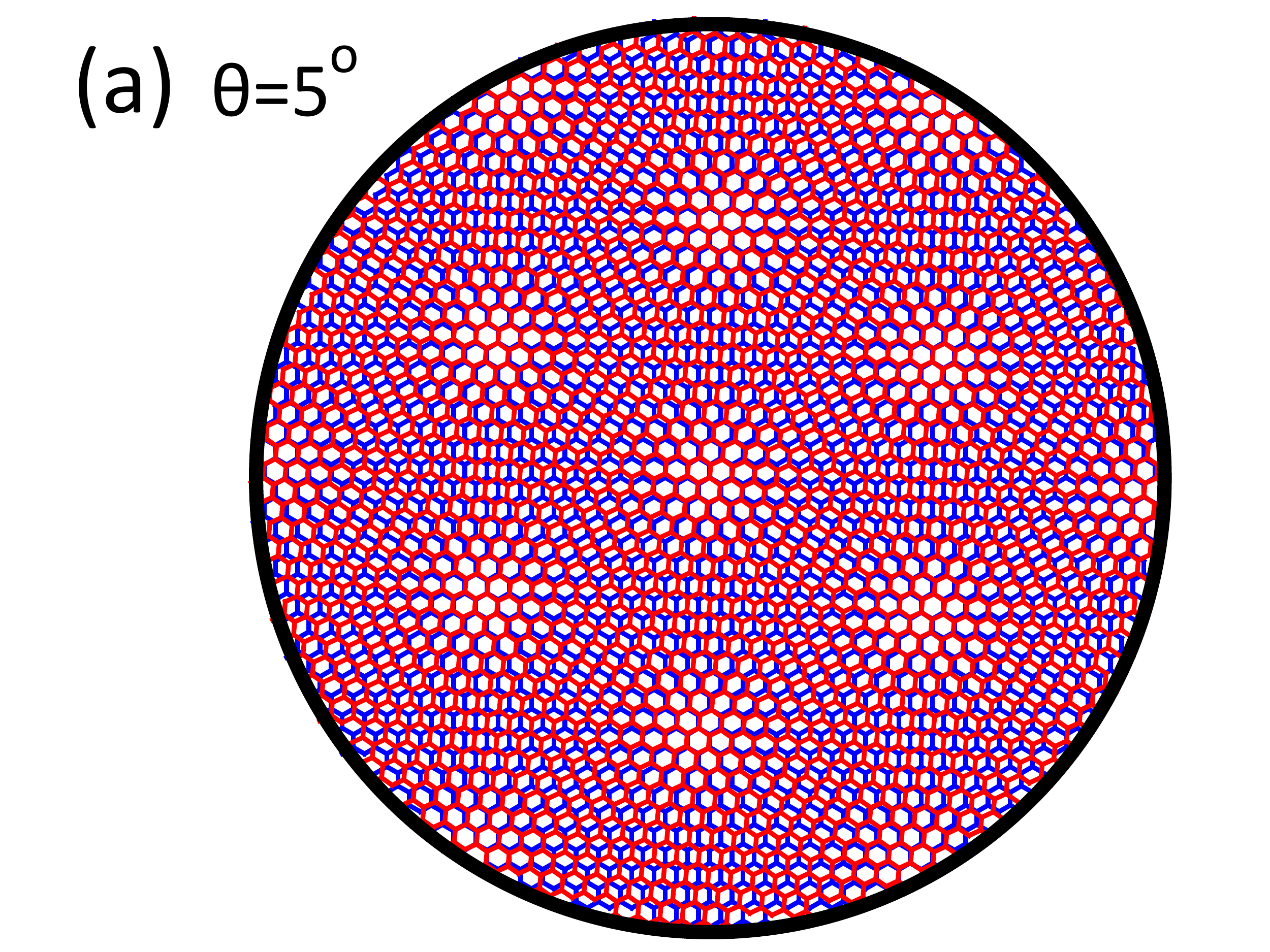}
\includegraphics[width=0.3\columnwidth]{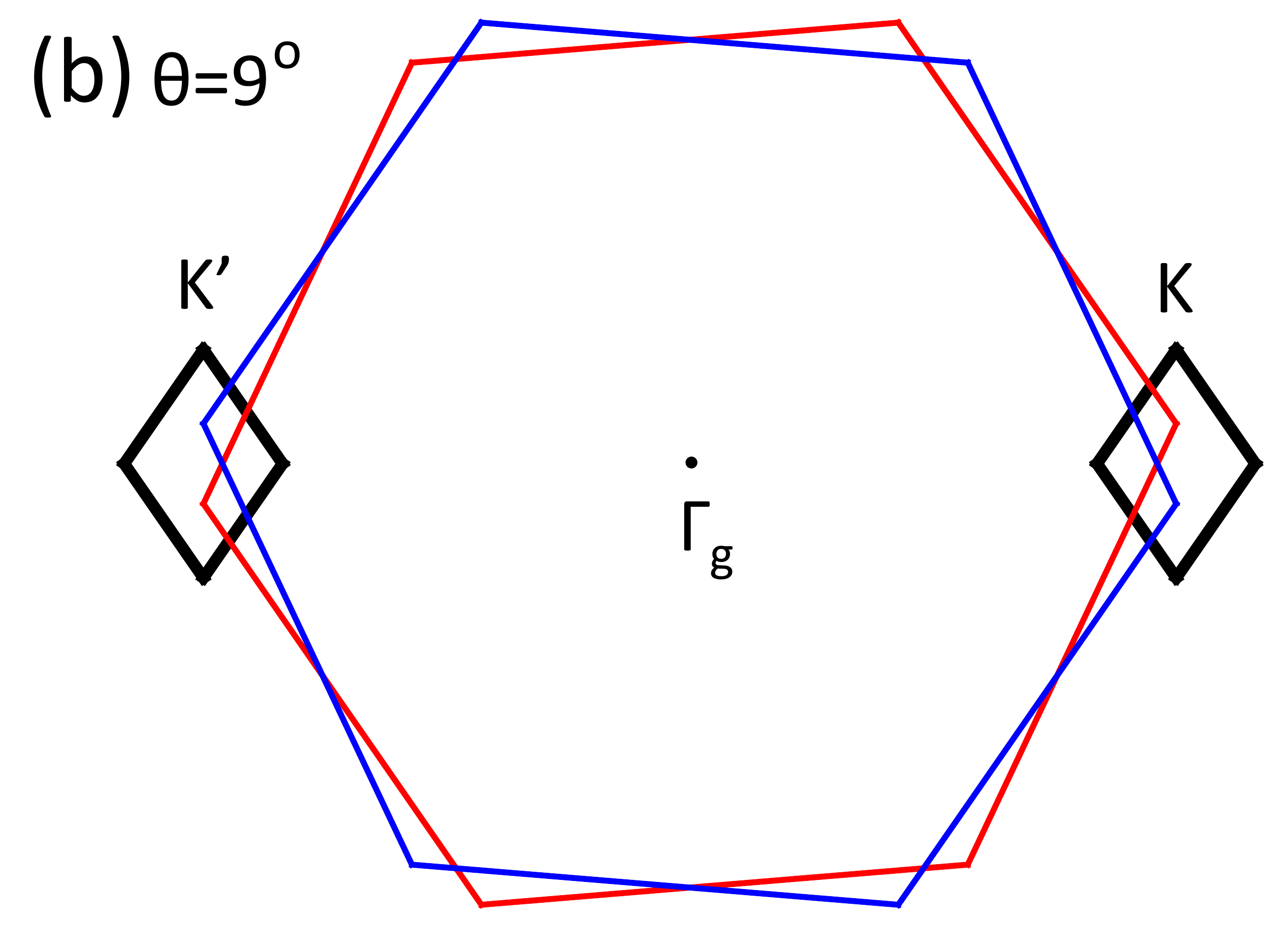}
\caption{(a) Real space image of TBG with a twist angle of $\theta=5^\circ$. (b) Single-layer Brillouin zone of graphene (blue and red curve) together with the two Moir\'e Brillouin zones (MBZ) of TBG belonging to each graphene $K$-valley (black curves) for $\theta=9^\circ$. 
\label{TBL}}
\end{figure*}

\section{Van Hove singularities in the highest valence band}

\subsection{Continuum model approach}

Here, we discuss the band-structure of TBG close to the neutrality point within the continuous model (CM) of Ref. \cite{Lopes07}. In Fig. \ref{BZvanHoveValley}, we show the band-structure of the highest valence band for one valley. The bands for the other valley are obtained by reflection. For TBG, the energy bands around the two non-equivalent $K$-points are thus not degenerate anymore (i.e., there is no parity which is the origin of the observed dichroism\cite{Kim16,Suarez17,Stauber18,Stauber18b}) but they are only related through time-reversal symmetry. The two highest valence bands $E_K$ and $E_{K'}$ belonging to valley $K$ and $K'$ have to be combined in order to be comparable to the bands obtained in the tight-binding calculation\cite{Suarez10,Trambly10,Moon13}. In the CM, there is no coupling between the different valleys and the two highest valence bands are, therefore, given by $E_+=\text{max}(E_{K},E_{K'})$ and $E_-=\text{min}(E_{K},E_{K'})$.

\begin{figure*}
\includegraphics[width=0.4\columnwidth]{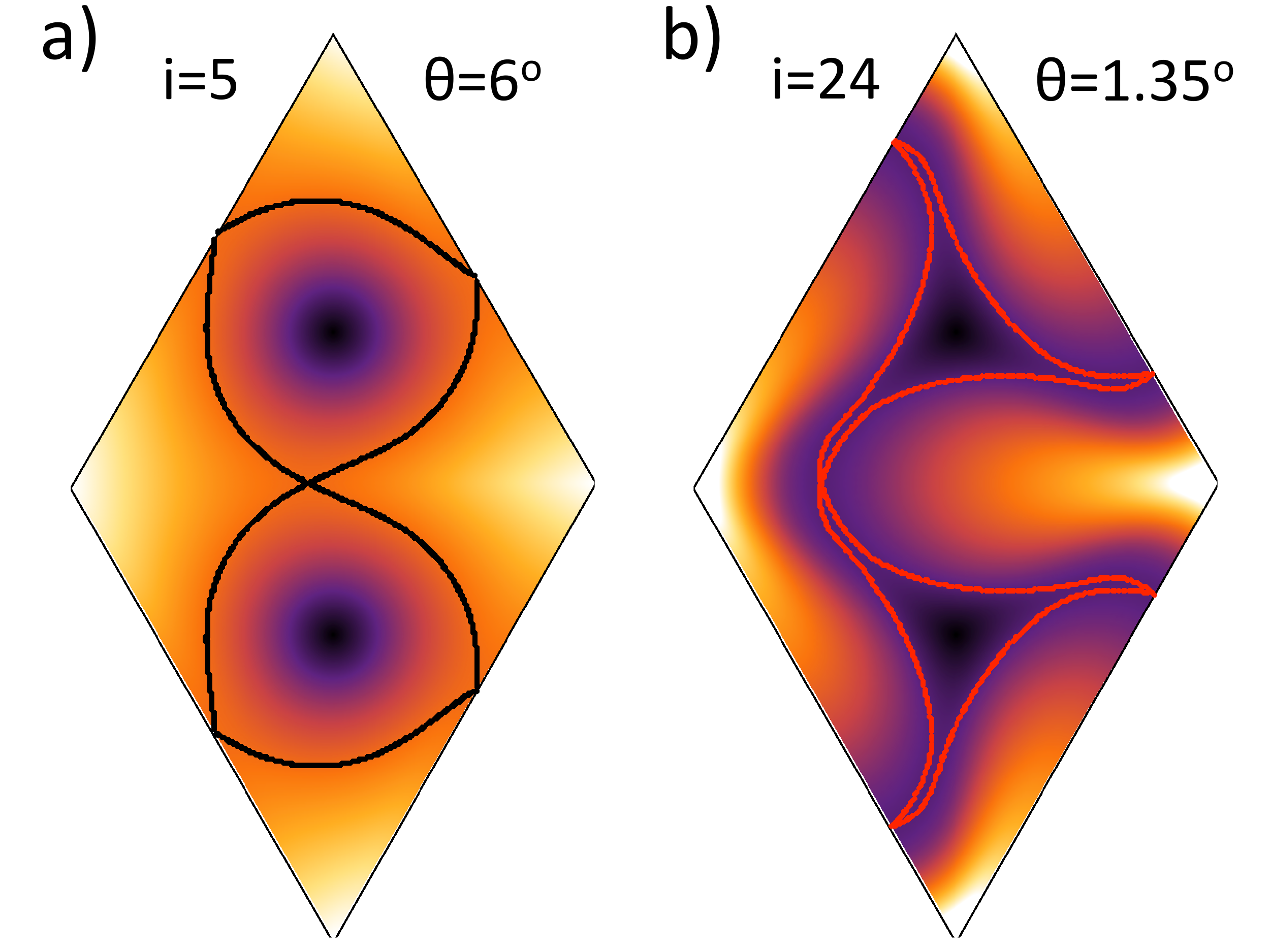}
\includegraphics[width=0.4\columnwidth]{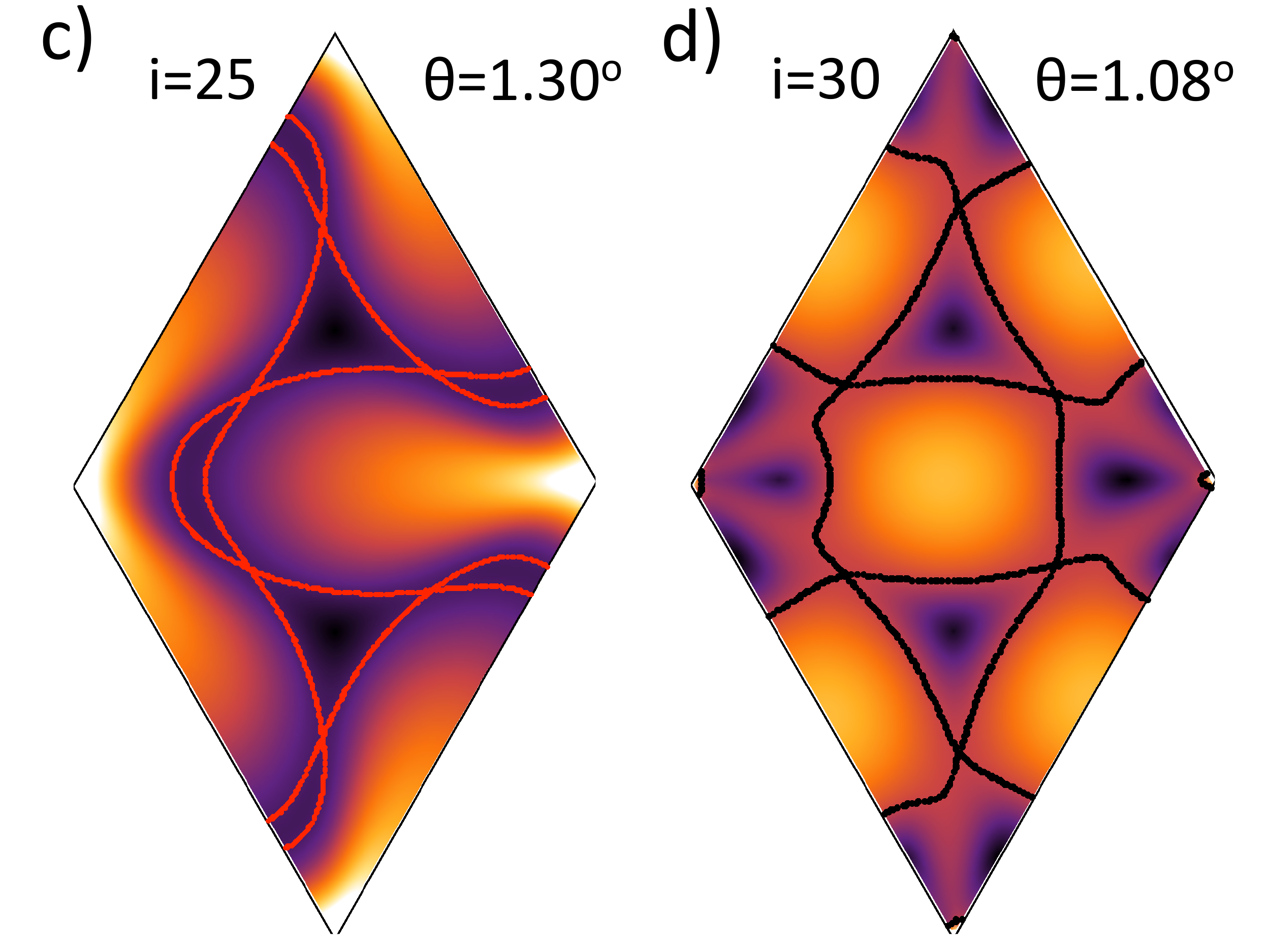}
\caption{Density plot of the energy dispersion of the highest valence band of TBG within the Moir\'e Brillouin zone (MBZ) around one valley, for four different twist angles. In all cases, dark (bright) colors represent high (low) energies and the $\Gamma$-point is located at the four corners of the rhombical MBZ around one valley. The black and red contour lines represent the Fermi surface at the energy of the van Hove singularity (vHS), $E_{vH}$, respectively. a) For large twist angle with $i\leq5$, the Fermi surface at $E_{vH}$ is given by a circular shape resembling the Dirac cone physics and the vHS (saddle points) are approximately located at the three $M$-points of the MBZ. b) For smaller twist angle with $5\leq i\leq24$, the three vHS slightly move away from the $M$-points and the Fermi surface at $E_{vH}$ becomes more and more triangular. c) At some critical angle $\theta_{i=24} > \theta_c^+ > \theta_{i=25}$, a doubling of the vHS occurs, i.e., there are now six vHS. d) For $25\leq i\leq30$, the six vHS are approximately located along the $\Gamma K_1$- and $\Gamma K_2$-directions, respectively. For a twist angle close to the magic angle with $i=30$, the Fermi surface at $E_{vH}$ is approximately built up of squares and triangles, with the Fermi level corresponding to a filling factor of $\approx 1/2$. 
\label{BZvanHoveValley}}
\end{figure*}

We will first look at the van Hove singularities (vHS) for the highest valence band $E_+$. For large twist angles, the Fermi surface at the van Hove energy $E_{\rm{vH}}$ can be approximated by a circle due to the isotropic Dirac cone physics expected for the highest valence band, see Fig. \ref{BZvanHovePlus} a). Decreasing the twist angle, the Fermi surface at $E_{\rm{vH}}$ becomes more an more triangular and the vHS (saddle points) move more and more away from the $M$-points. For $i\approx 24-25$, we observe a splitting of the vHS, see Fig. \ref{BZvanHovePlus} b) and c). The exact crossing point usually occurs at a non-commensurate critical angle $\theta_c^+$ that can be treated by more advanced numerical techniques\cite{Carr17,Cances17}. 

\begin{figure*}
\includegraphics[width=0.4\columnwidth]{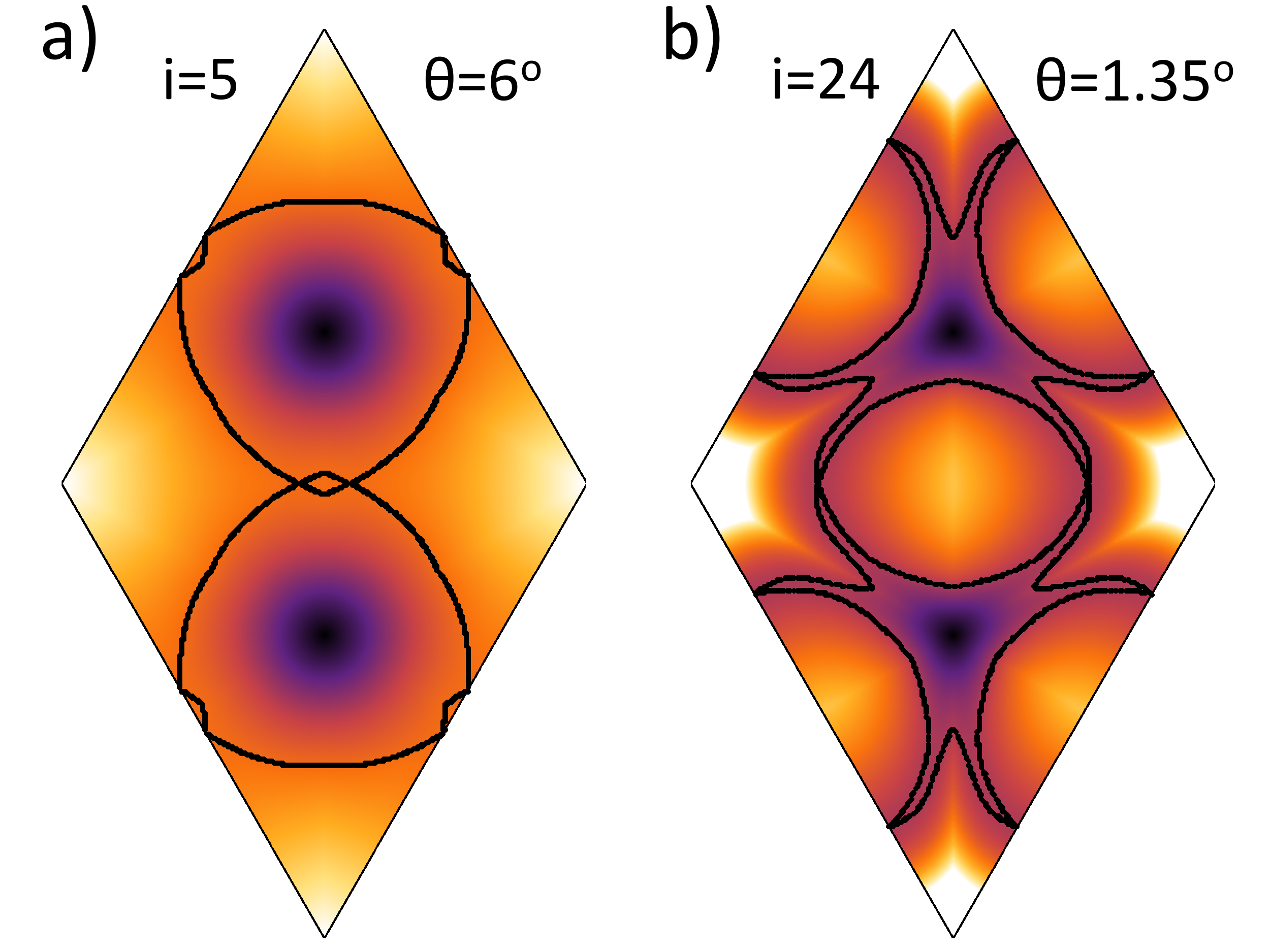}
\includegraphics[width=0.4\columnwidth]{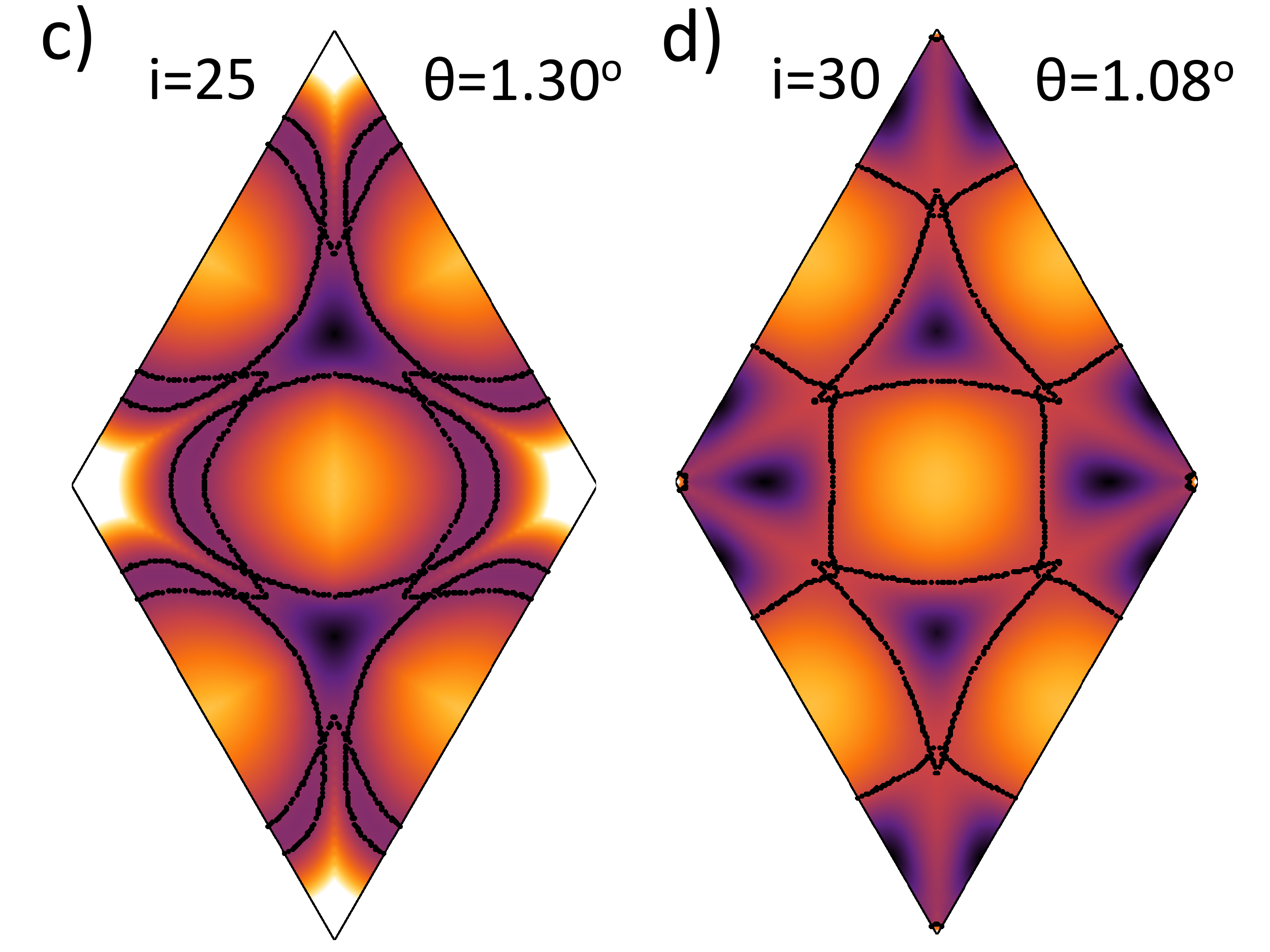}
\caption{Density plot of the energy dispersion of the highest valence band $E_+$ of TBG on the MBZ including both valleys, for four different twist angles. In all cases, dark (bright) colors represent high (low) energies and the $\Gamma$-point is located at the four corners of the rhombic MBZ. The black contour lines represent the Fermi surface at the energy of the vHS, $E_{\rm{vH}}$. Panel a)-d) shows the same evolution as in the case of one valley shown in Fig. \ref{BZvanHoveValley}. a) For large twist angle with $i\leq5$, the Fermi surface at $E_{\rm{vH}}$ is given by a circular shape resembling the Dirac cone physics and the vHS are symmetrically located around the three $M$-points of the MBZ. b) For smaller twist angle with $5\leq i\leq24$, the Fermi surface at $E_{\rm{vH}}$ becomes more triangular and the six vHS move further away from the $M$-points, but remain on the center and zone boundary of the MBZ. c) At some critical angle $\theta_{i=24} > \theta_c^+ > \theta_{i=25}$, a doubling of the vHS occurs, i.e., there are now twelve vHS located inside the MBZ close to the lines that connect the $\Gamma$ and  $K_\ell$-points. d) For a twist angle close to the magic angle with $i\leq30$, the twelve vHS move to the midpoint of the $\Gamma K_\ell$-line and the Fermi surface at $E_{\rm{vH}}$ is approximately built up of hexagons, squares and triangles, with the Fermi level corresponding to a filling factor of $\approx1/2$. 
\label{BZvanHovePlus}}
\end{figure*}

Also for the second highest valence band $E_-$, a critical angle $\theta_c^-$ can be defined that marks a change in the topology of the band. For large twist angles, the vHS are precisely located at the three $M$-points and, for larger energies, no additional singularities exist. But for $i\approx26-27$, new vHS develop along the $\Gamma K_\ell$-directions, see Fig. \ref{BZvanHoveMinus} b) and c). And for  $i\approx30$, they are located at the midpoint between the $\Gamma$ and $K_\ell$-points, see Fig. \ref{BZvanHoveMinus} d).

\begin{figure*}
\includegraphics[width=0.4\columnwidth]{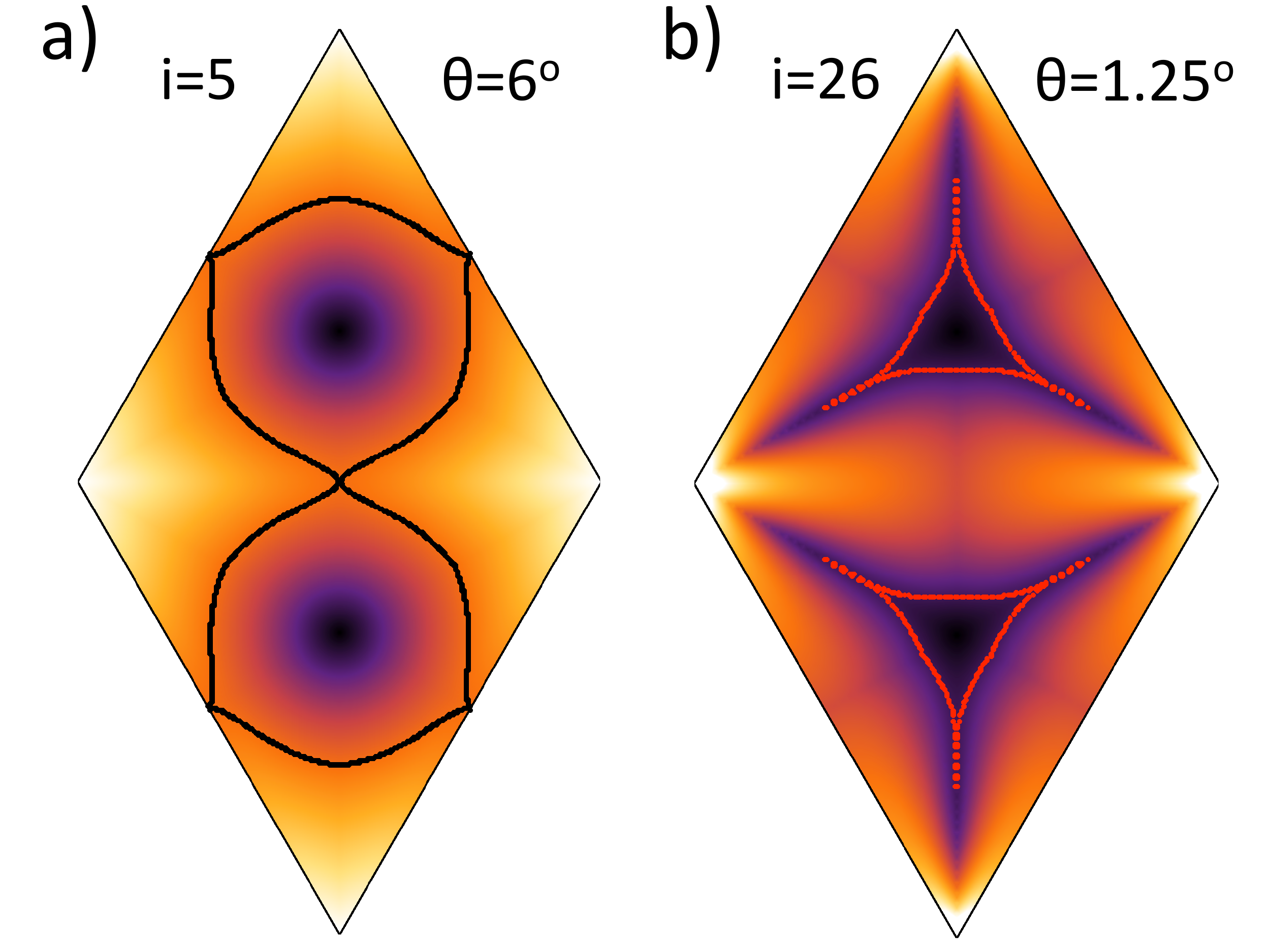}
\includegraphics[width=0.4\columnwidth]{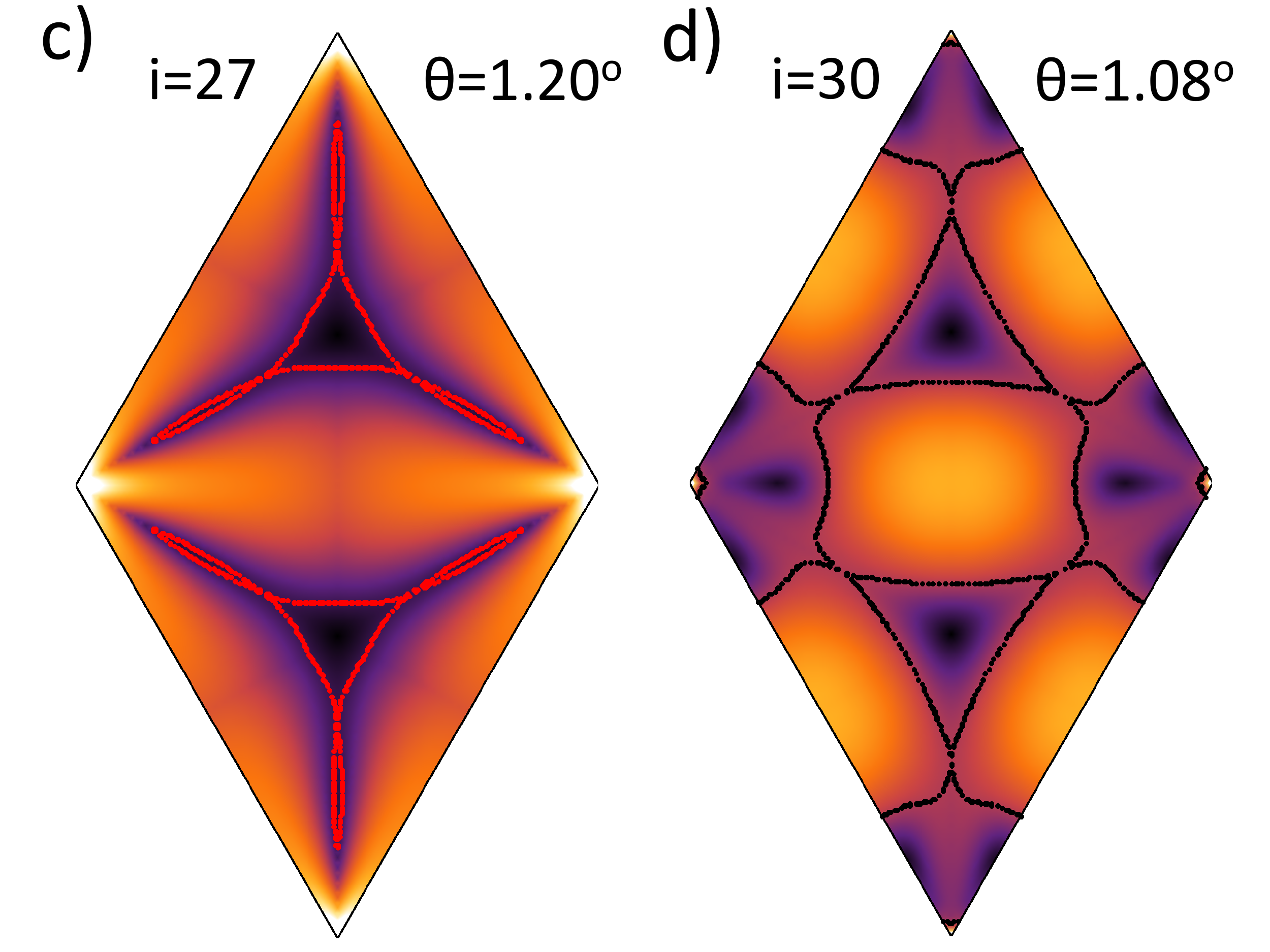}
\caption{Density plot of the energy dispersion of the second highest valence band $E_-$ of TBG on the MBZ including both valleys, for four different twist angles. In all cases, dark (bright) colors represent high (low) energies and the $\Gamma$-point is located at the four corners of the rhombic MBZ. The black and red contour lines represent the Fermi surface at or close to the energy of the vHS. a) For large twist angles, the Fermi surface is approximately circular and the vHS are located at the $M$-points. b) For a twist angle with $i=26$, the Fermi surface at a certain energy is close to develop a vHS. c) For the twist angle with $i=27$, new vHS have developed and are located at the lines that connect the $\Gamma$ and $K_\ell$-points. d) For a twist angle with $i=30$, the vHS move to the midpoint of the $\Gamma K_\ell$-lines. \label{BZvanHoveMinus}}
\end{figure*}

The critical angle of the highest valence band $\theta_c^+\approx1.33^\circ$ is slightly larger than the critical angle of the second highest valence band $\theta_c^-\approx1.23^\circ$, but the locations of the twelve vHS in the $E_+$-bands are practically identical to the locations of the six vHS in the $E_-$-bands. Still, we could have chosen a different representation of the eigenenergies by only considering the energy bands of the two independent valleys, separately. Then, we would have found six vHS in the $E_K$-bands and six vHS in the $E_{K'}$-bands, all at the same energy $E_{\rm{vH}}$. The vHS in the $E_-$-bands are thus the result of the "folding" of the two $E_K$ and $E_{K'}$-bands which are degenerate along the $\Gamma K_\ell$-lines. This means that a small but finite coupling between the two valleys, naturally present for instance in a tight-binding scheme, may be relevant to obtain a more accurate shape of the highest valence bands near the $\Gamma K_\ell$-lines.

\subsection{Tight-binding approach}

We present next the evolution of the saddle points obtained from the tight-binding model (TBM). A main difference with respect to the above CM description is that the valley index is not conserved in the TBM, so there is always an intrinsic coupling between different valleys. This is usually negligible but, due to the small bandwidths of the energy bands around the charge neutrality point and the degeneracy of the energy bands of the two valleys along the high-symmetry lines of the MBZ, a small valley coupling may result in significant changes in the band topology.

Fig. \ref{BandStructure} shows for instance the two highest valence and two lowest conduction bands obtained from the Hamiltonian (\ref{tbh}), for different twisted bilayers from large to small twist angle. For the TBM used here with no bound in the interlayer hopping range, there appears a close degeneracy between the two valence bands (and the two conduction bands) along the high-symmetry line $\Gamma KM$, as can be seen in the plots. Anyhow, we have to keep in mind that the curvatures of the two valence bands (as well as of the two conduction bands) are different away from the high-symmetry line, leading to different topologies which can be more clearly appreciated in the contour plots shown below.

\begin{figure*}
\includegraphics[width=0.24\columnwidth]{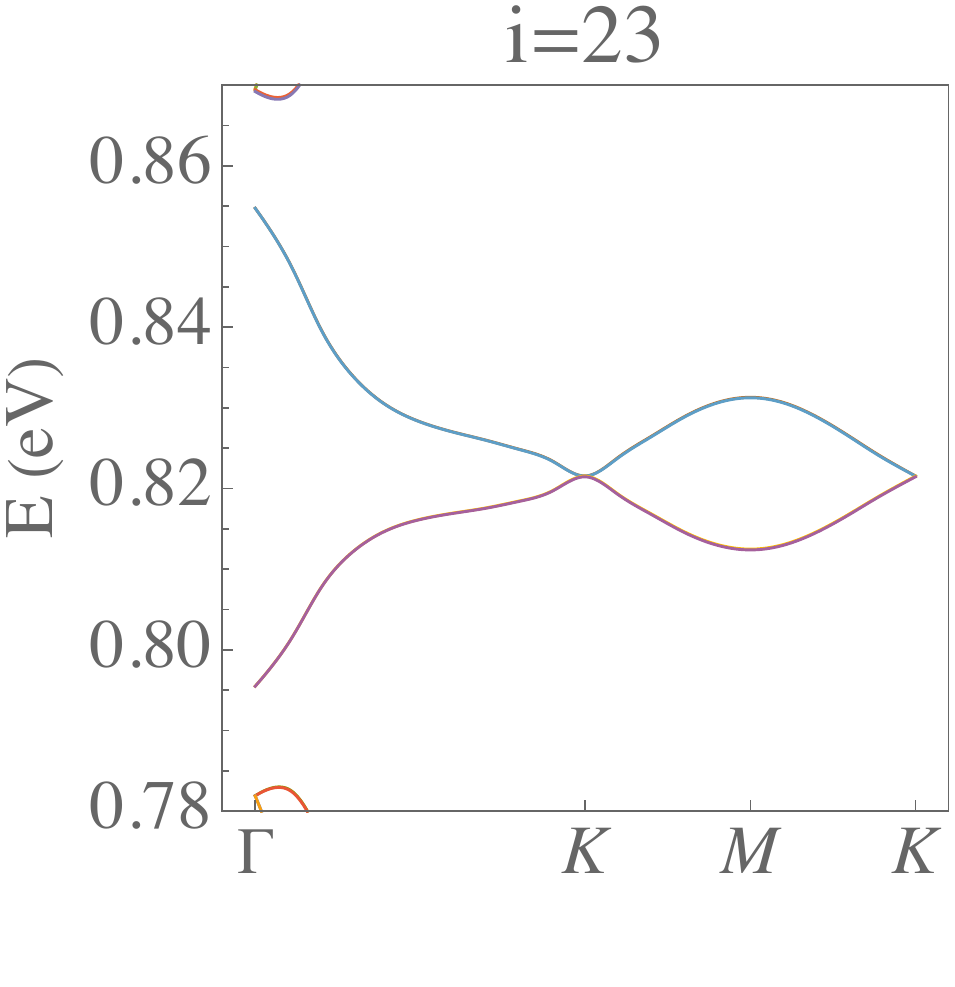}
\includegraphics[width=0.24\columnwidth]{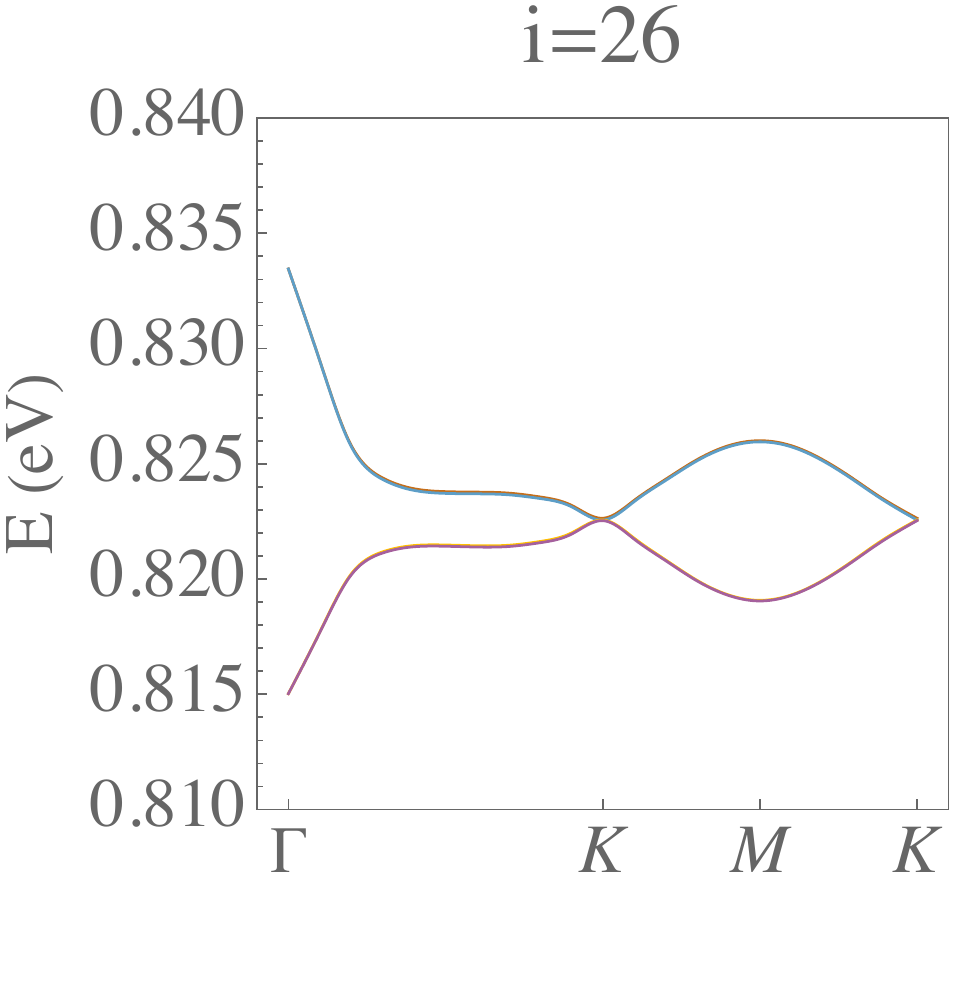}
\includegraphics[width=0.24\columnwidth]{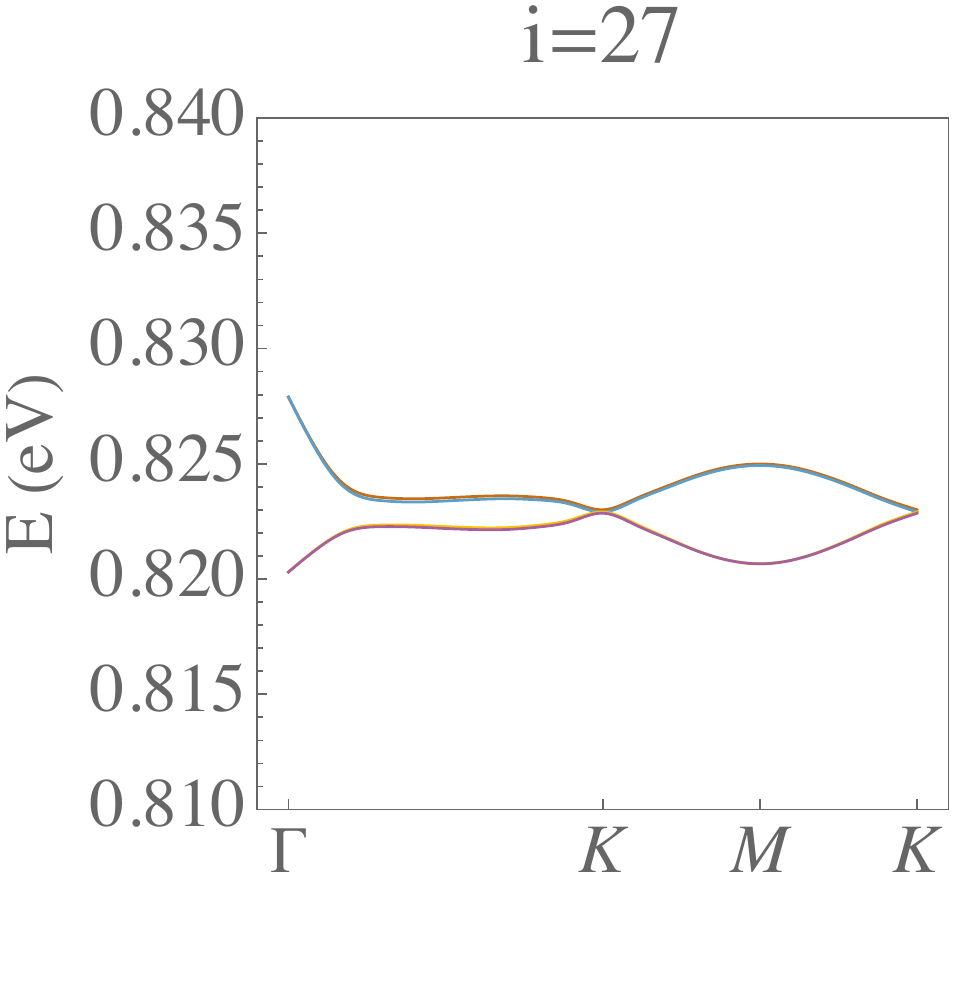}
\includegraphics[width=0.24\columnwidth]{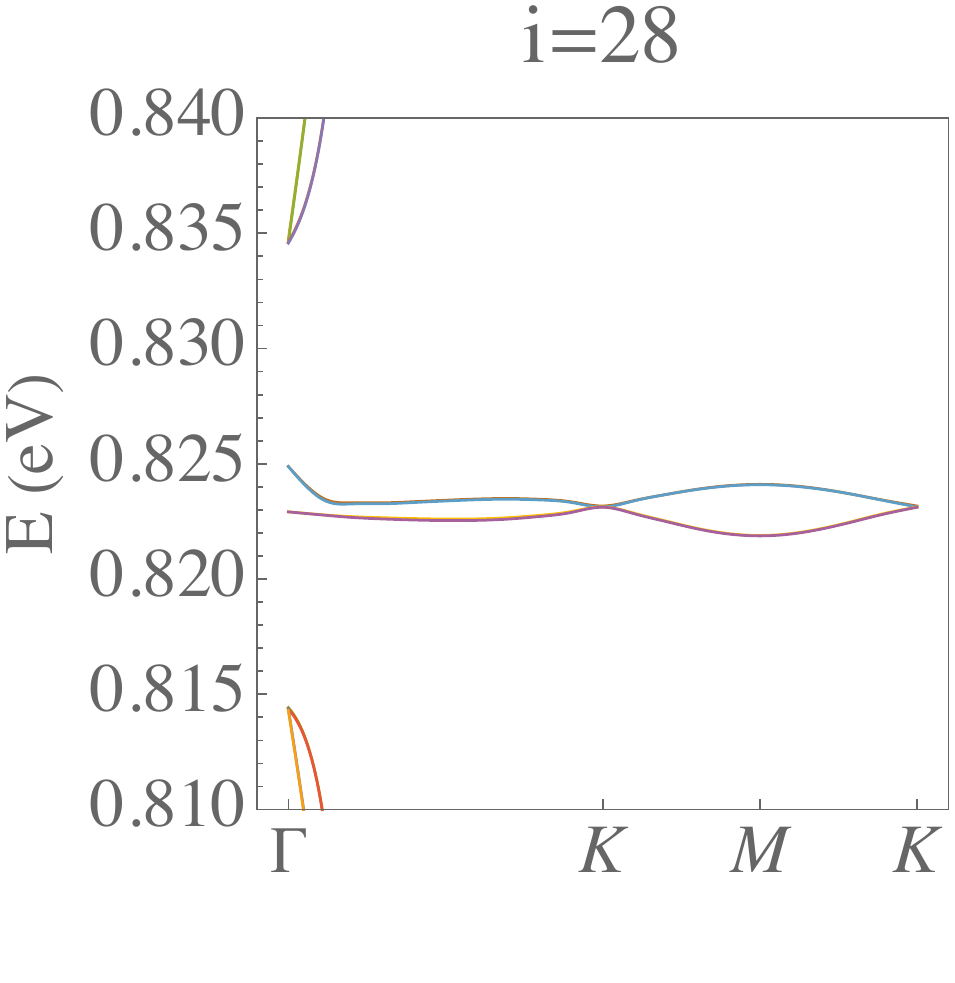}
\caption{Band structure along the high-symmetry line $\Gamma KM$ of the two highest valence and two lowest conduction bands of the TBM for four different twist angles progressively closer to the magic angle. 
\label{BandStructure}}
\end{figure*}

In the tight-binding approach, we can also focus on the evolution of the saddle points in the two highest valence bands, which show a behavior very similar to that already found with the CM. Thus, for sufficiently large twist angles down to $\approx 1.35^\circ$, we find that the saddle points in the valence band $E_-$ are always pinned at the $M$ point. On the other hand, there are also saddle points in the highest valence band $E_+$ placed along the $\Gamma M$ line, which move progressively away from the $M$ point as the twist angle decreases. This is illustrated in the contour plot shown in Fig. \ref{tbj3} a), which represents the energy contour map in the tight-binding approach for the highest valence band of the twisted bilayer with $i = 23$. 

Furthermore, the tight-binding approach shows that there is a critical twist angle, corresponding to the critical point $\theta_c^+$ found in the CM, where each saddle point in the valence band $E_+$ splits in a pair of saddle points which move away from the $\Gamma M$ line for decreasing twist angle. In complete analogy with the behavior already found in the CM, there is then a doubling in the number of saddle points in the highest valence band $E_+$, with the new saddle points in each pair moving progressively towards the $\Gamma K$ line as the twist angle is lowered. This can be appreciated in Fig. \ref{tbj3} b), which represents the energy contour map of the highest valence band $E_+$ for the twisted bilayer with $i = 25$. The plot shows a quite similar topology to that of the band $E_+$ obtained in the CM, as can be observed from the comparison with Fig. \ref{BZvanHovePlus} c) above.

\begin{figure*}
\includegraphics[width=0.2\columnwidth]{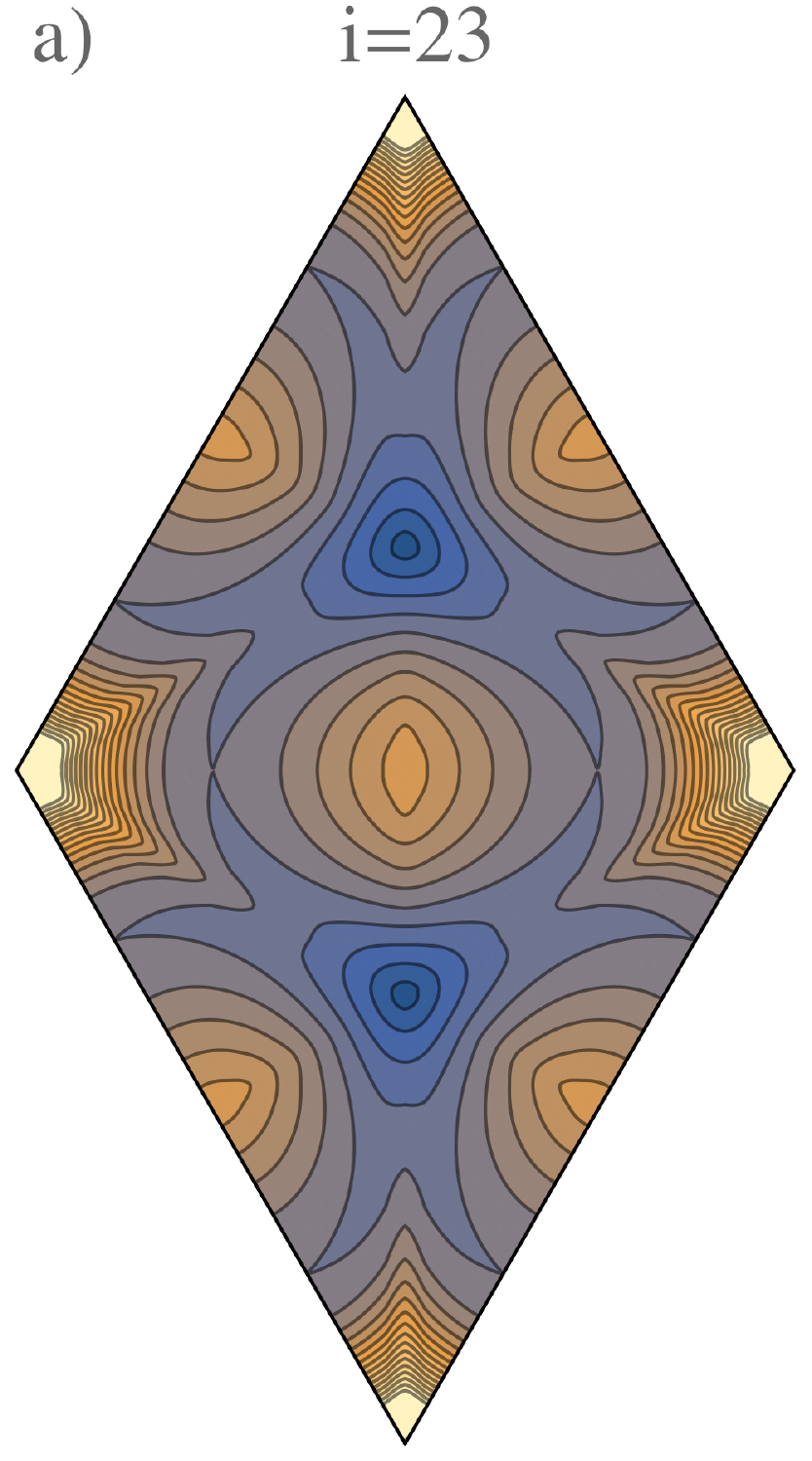}
\includegraphics[width=0.2\columnwidth]{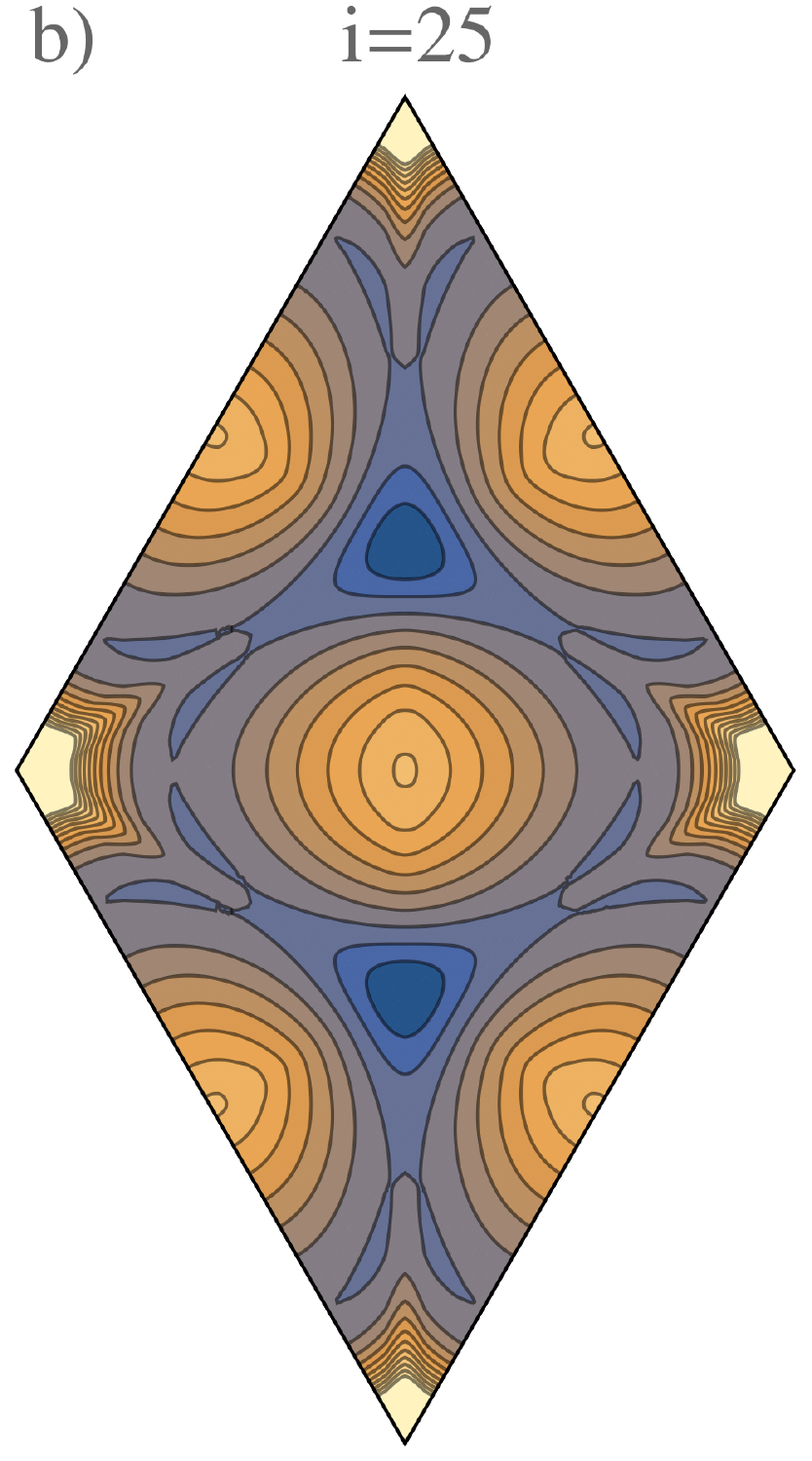}
\includegraphics[width=0.2\columnwidth]{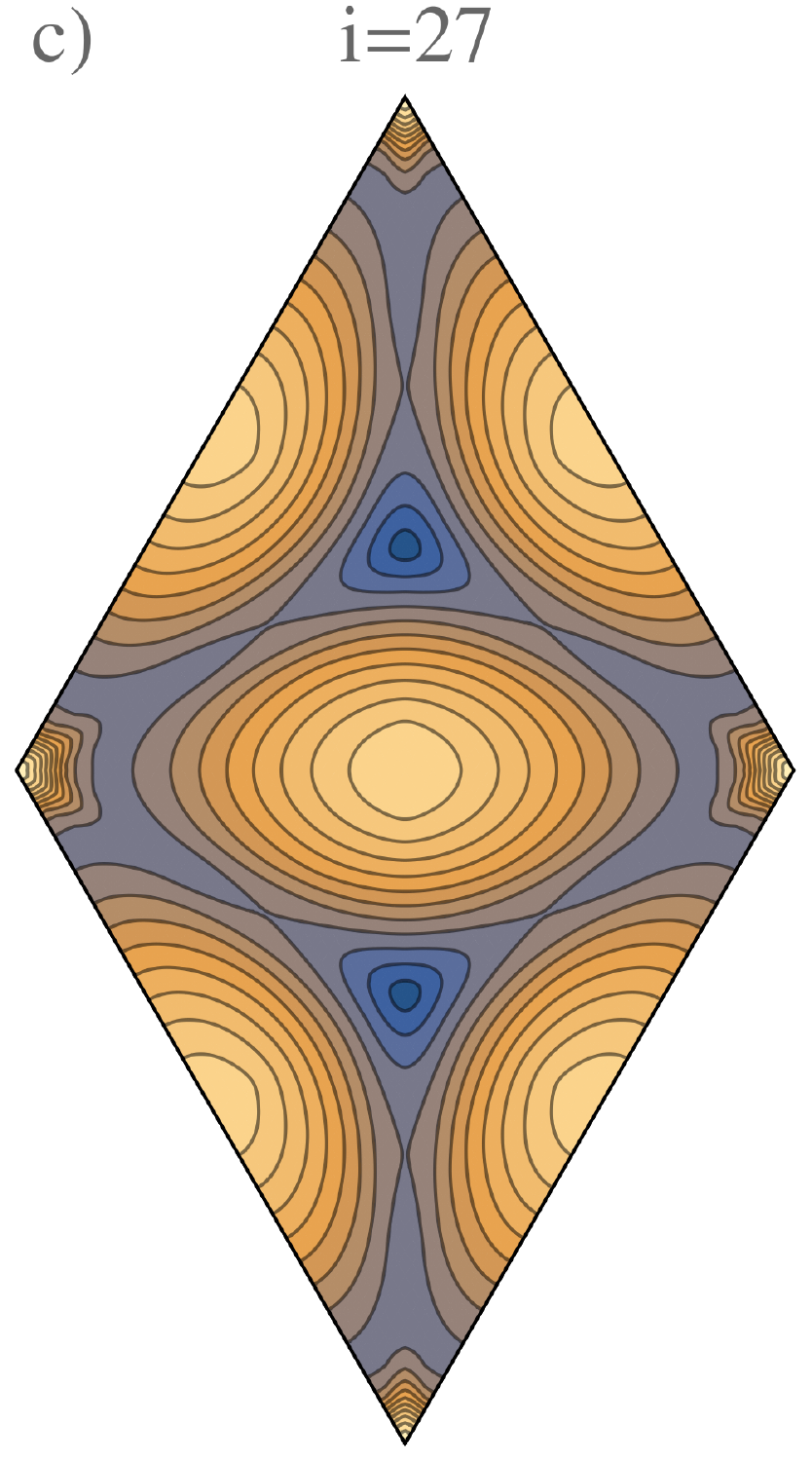}
\includegraphics[width=0.2\columnwidth]{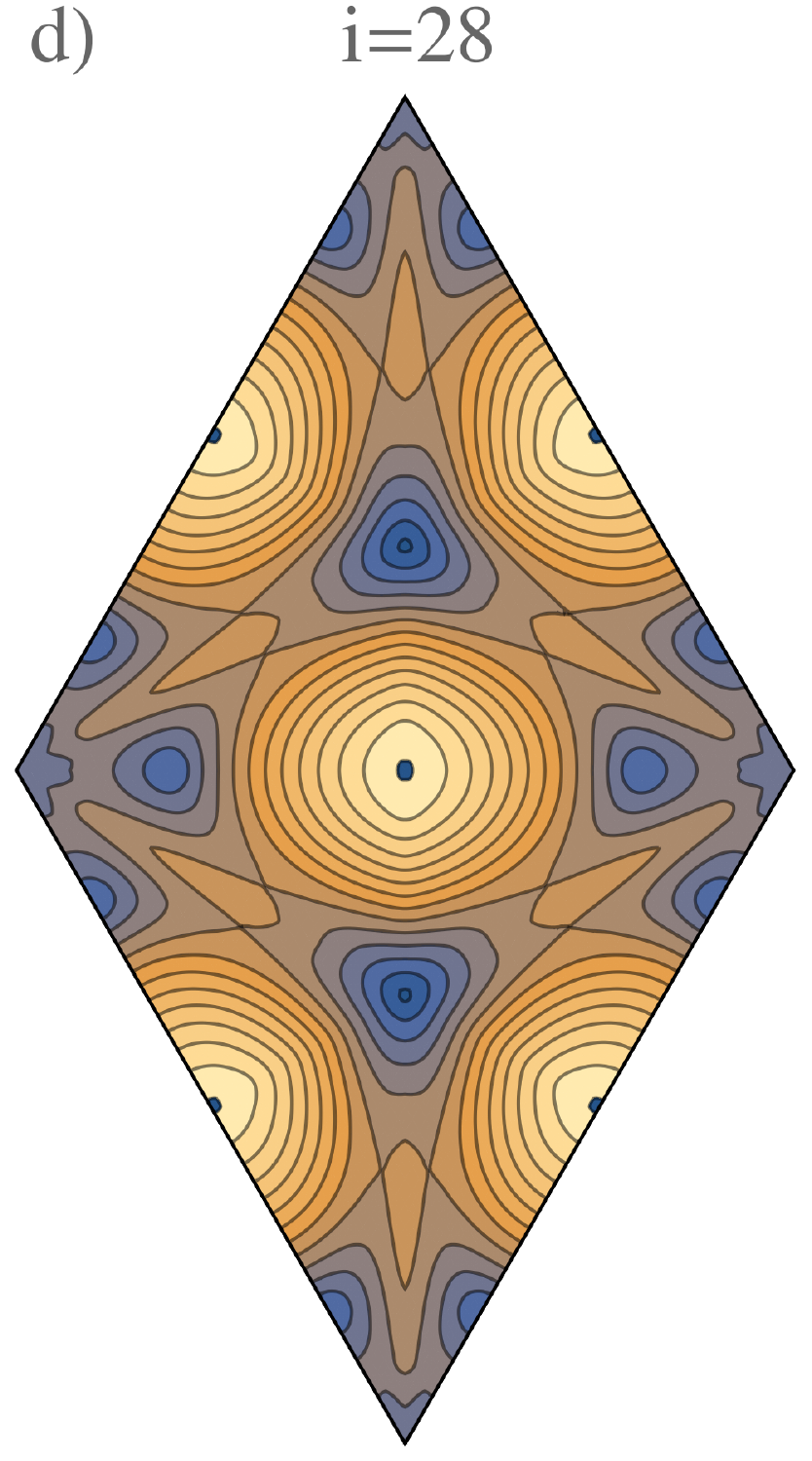}
\caption{Energy contour maps of the highest valence band $E_+$ in the MBZ of TBG for four different twist angles. In all cases, dark (bright) colors represent high (low) energies. The $\Gamma$-point is located at the four corners of the rhombic MBZ, with each side corresponding to high-symmetry lines $\Gamma M \Gamma$. The $K$-points coincide with the peaks with highest energy at the interior of the MBZ.}
\label{tbj3}
\end{figure*}

Decreasing further the twist angle, we find that for the twisted bilayer with $i = 26$ the saddle points that were approaching the $\Gamma K$ line have already merged in pairs over that line. This confirms the existence of another critical point which corresponds to the critical angle $\theta_c^-$ already found in the CM. At this critical point, a drastic change in the topology of the highest valence bands takes place, as we observe that there is a transfer of a saddle point in each pair from the highest valence band $E_+$ to $E_-$. This becomes clear from inspection of the contour plots in Fig. \ref{tbj4}, which shows the evolution in the tight-binding approach of the energy contour maps of the band $E_-$ for twisted bilayers from $i = 23$ to $i = 28$.

\begin{figure*}
\includegraphics[width=0.2\columnwidth]{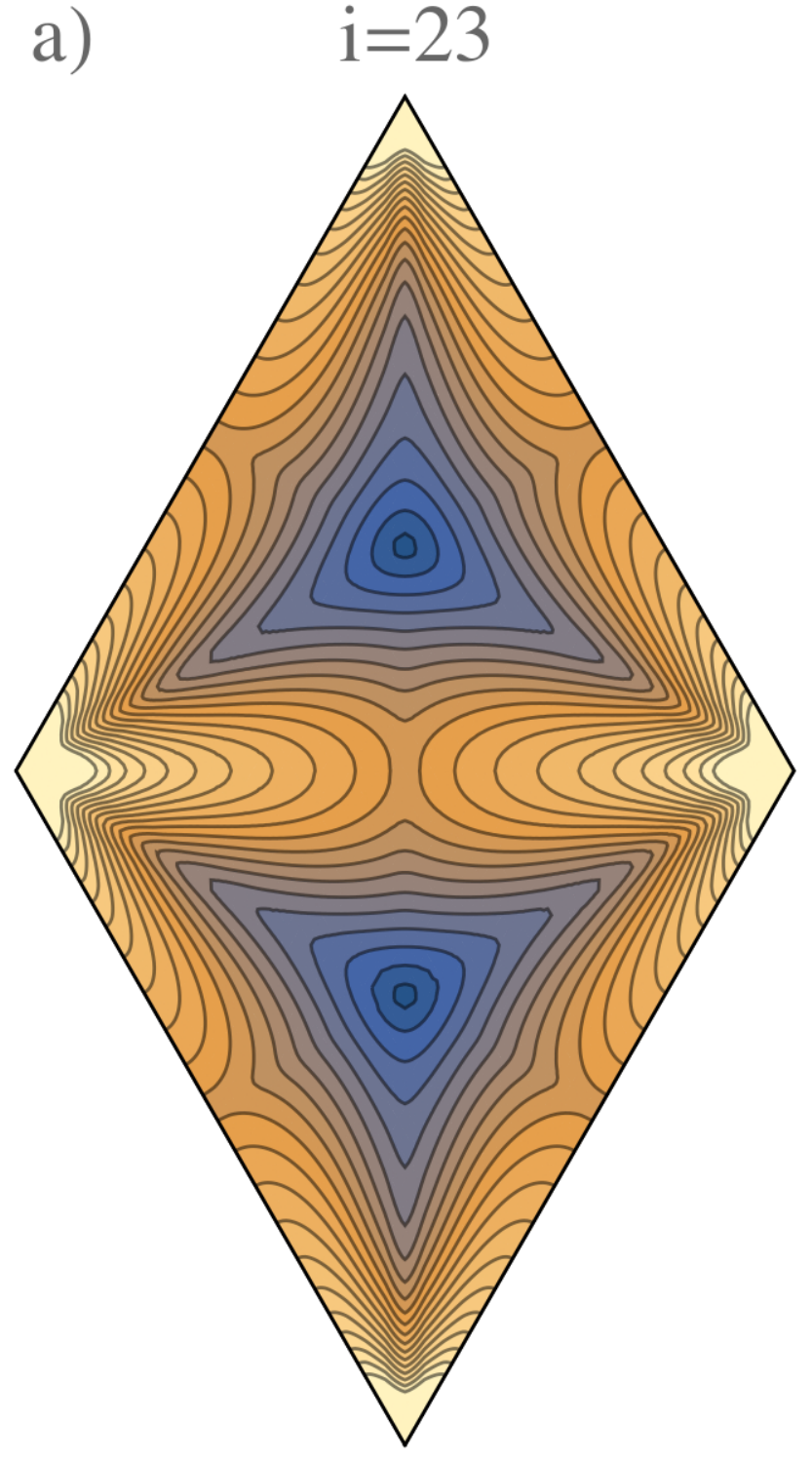}
\includegraphics[width=0.2\columnwidth]{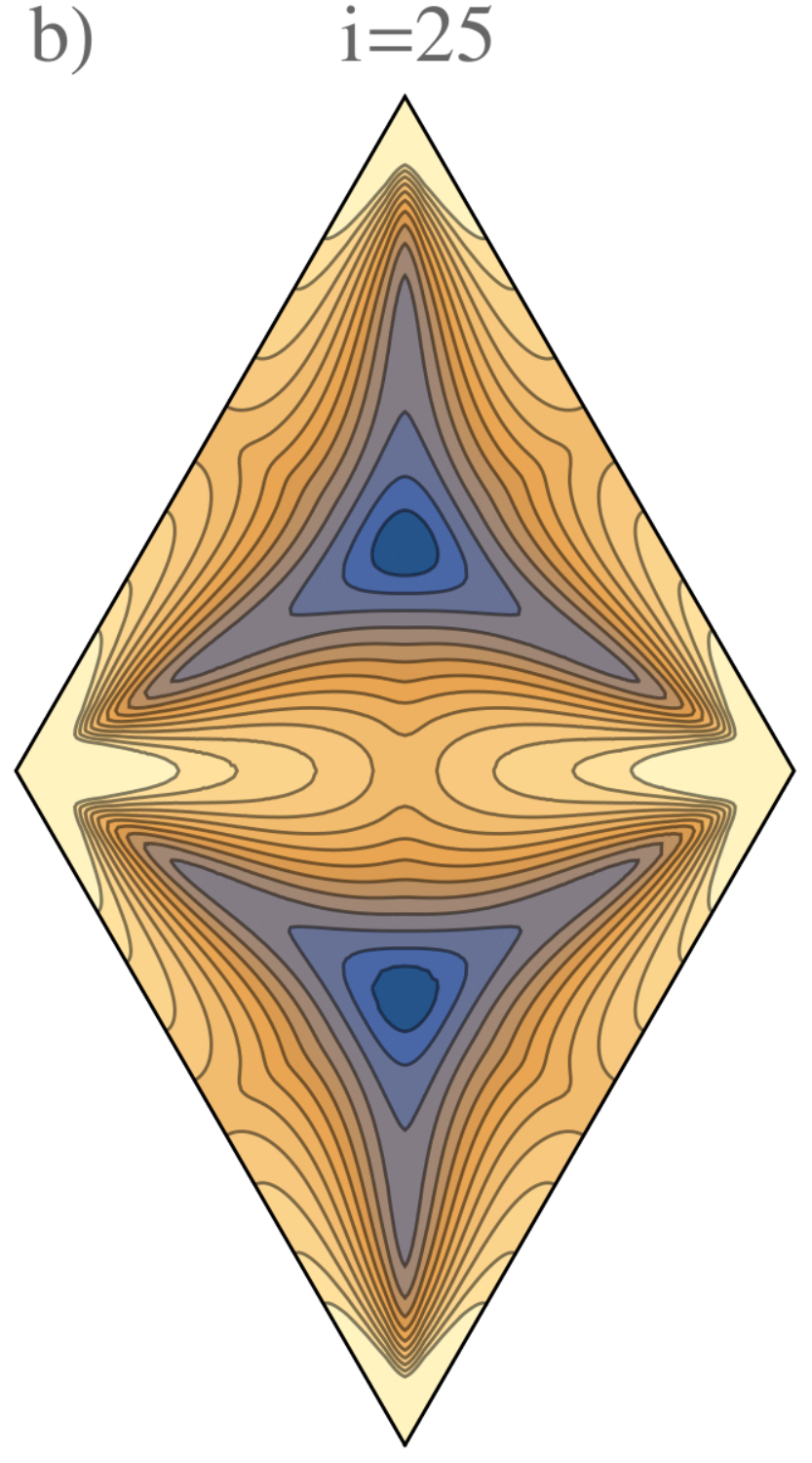}
\includegraphics[width=0.2\columnwidth]{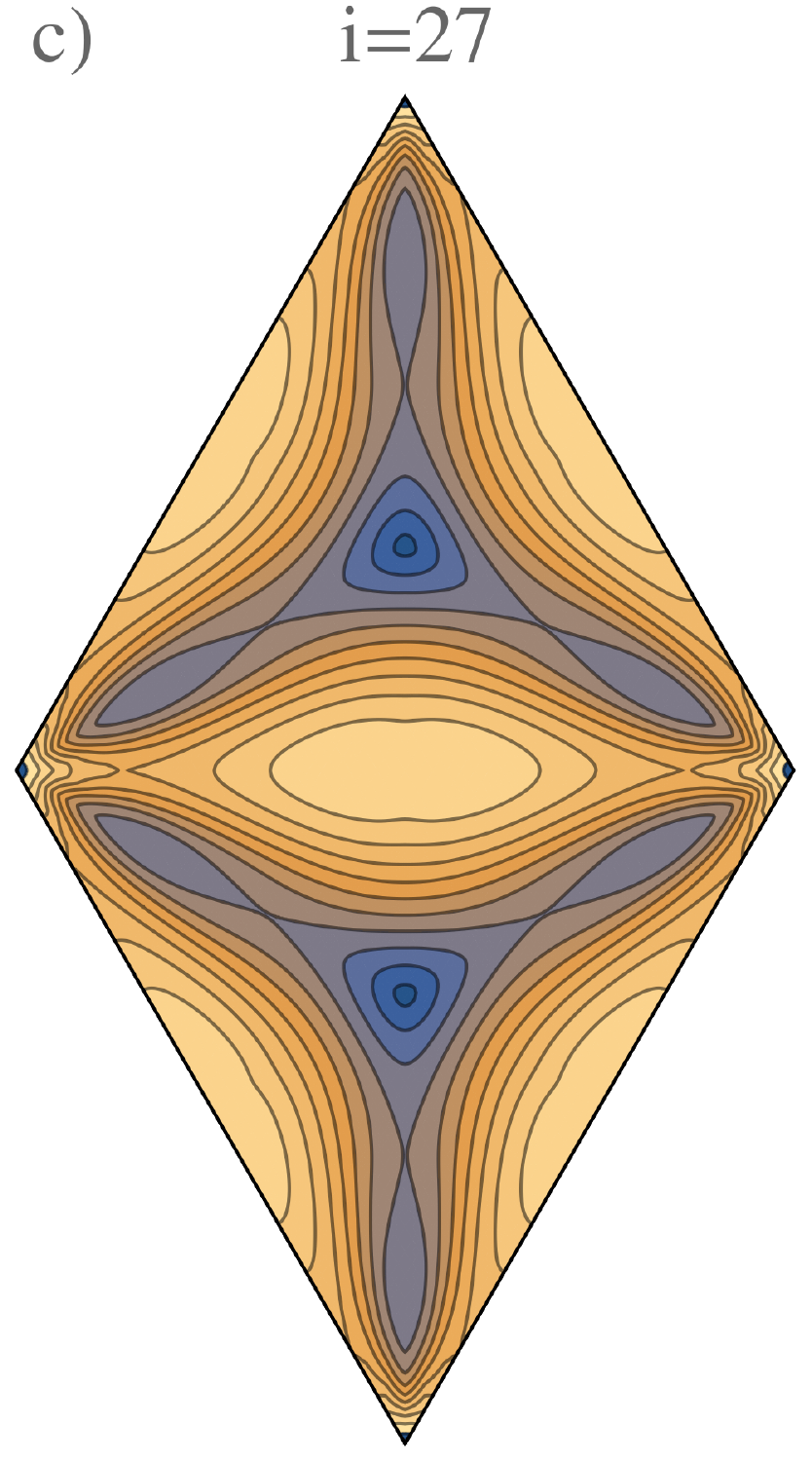}
\includegraphics[width=0.2\columnwidth]{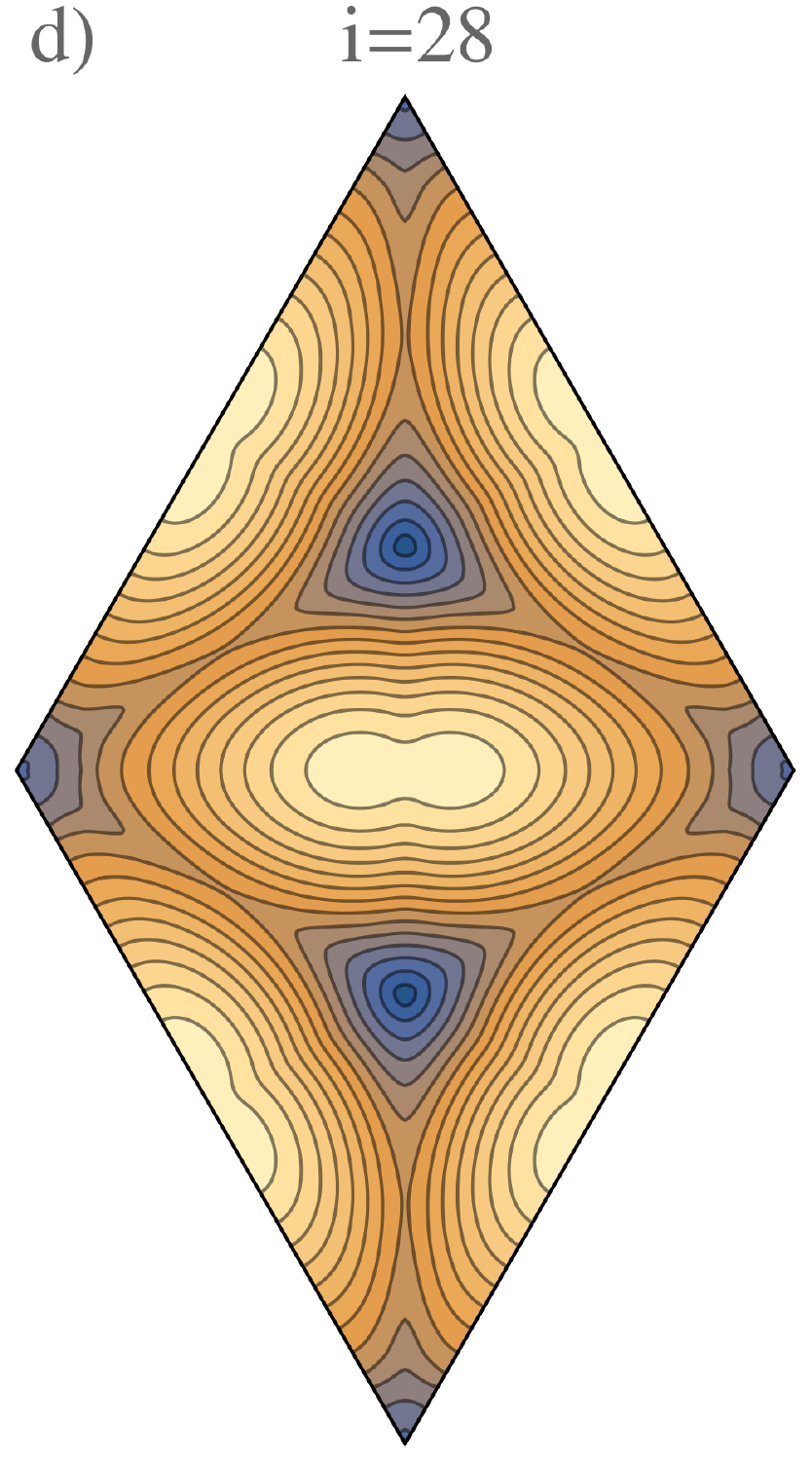}
\caption{Energy contour maps of the second highest valence band $E_-$ in the MBZ of TBG for four different twist angles. In all cases, dark (bright) colors represent high (low) energies. The $\Gamma$-point is located at the four corners of the rhombic MBZ, with each side corresponding to high-symmetry lines $\Gamma M \Gamma$. The $K$-points coincide with the peaks with highest energy at the interior of the MBZ.}
\label{tbj4}
\end{figure*}

We find then a very close agreement between the description of the twisted bilayers carried out within the CM approach and the tight-binding approach. We arrive at a very consistent picture of the evolution of the saddle points in the highest valence bands $E_+$ and $E_-$, unveiling the existence of two critical twist angles $\theta_c^+$ and $\theta_c^-$. These lead to universal features in the highest valence bands of the twisted bilayers, as they are found in quite complementary approaches based respectively in the CM and the TBM. That universal character is also supported by the fact that the two critical points correspond to changes in the topology of the bands, which has to be insensitive to small perturbations of the twisted bilayers. 
This supports the robustness of the properties described in the main text for the twisted bilayer with $i = 26$, and which should extend to bilayers with smaller twist angle as long as they are rooted in features linked to the topology, like the extended character of the saddle points and the approximate nesting of the energy contour lines near the vHS of the valence band $E_-$.      

\subsubsection{Tight-binding approach with lattice relaxation}
\label{Universality}
The proposed Kohn-Luttinger mechanism crucially depends on the doubling of the van Hove singularity. Here, we will thus address the universality of this feature by calculating the band-structure of the tight-binding Hamiltonian considering also in-plane lattice relaxation which distorts the lattice especially for small twist angles. In the subsequent analysis, we will follow the procedure outlined in Ref. \cite{Nam17}, including the lattice relaxation within a continuous elasticity theory.

Our results are shown in Fig. \ref{vanHoveUniversality} for the commensurate twist angle corresponding to $i=22, 27$ and $i=31, 60$. In the first two cases, the vHS lie on the zone boundary of the rhomical Brillouin zone, whereas for $i=31$ and $i=60$, the vHS lie inside the Brillouin zone, i.e., a doubling of the vHS has taken place. This shows that the doubling of the vHS is seen for different realisations of the tight-binding model including such strong perturbation as lattice relaxation. Nevertheless, a detailed study within an improved model for the lattice relaxation is left for future studies.  

\section{Cooper-pair scattering} 
The microscopic interaction of the generalised BCS theory including triplet pairing only includes the scattering between electron pairs with vanishing momentum. Denoting the Cooper pairs as $c_\k^{\sigma,\sigma'}=a_{\k,\sigma}a_{-\k,\sigma'}$, we can write the corresponding Hamiltonian as:
\begin{align}
H=\sum_{\k,\sigma}\xi_\k a_{\k,\sigma}^\dagger a_{\k,\sigma}+\frac{1}{2}\sum_{\k,\k';\tau,\tau',\sigma,\sigma'} V_{\k,\k'}^{\tau,\tau';\sigma,\sigma'}\left(c_\k^{\tau,\tau'}\right)^\dagger c_{\k'}^{\sigma,\sigma'}
\end{align}
Projecting onto the non-interacting ground-state with $n_{\k,\sigma}=\langle a_{\k,\sigma}^\dagger a_{\k,\sigma}\rangle$, we obtain 
\begin{align}
\langle H\rangle=\sum_{\k,\sigma}\xi_\k n_{\k,\sigma}+\frac{1}{2}\sum_{\k;\sigma,\sigma'} \left(V_{\k,\k}^{\sigma,\sigma';\sigma,\sigma'}-V_{-\k,\k}^{\sigma',\sigma;\sigma,\sigma'}\right) n_{\k,\sigma}n_{-\k,\sigma'}
\end{align}
The Hartree term is thus defined by spin-singlet scattering with $\k-\k'=0$, whereas the Fock term is given by triplet scattering with $\k+\k'=0$. It also becomes obvious that the exchange term can lead to an attractive channel. This argument holds independent of the patch structure of the Fermi line. We thus expect attractive channels also in the case of small twist angles close to the magic angle (e.g. $i=30$ in the CM) where the Fermi line consists of three patches, see Figs. \ref{BZvanHovePlus} d) and \ref{BZvanHoveMinus} d).

\begin{figure*}
\includegraphics[width=0.49\columnwidth]{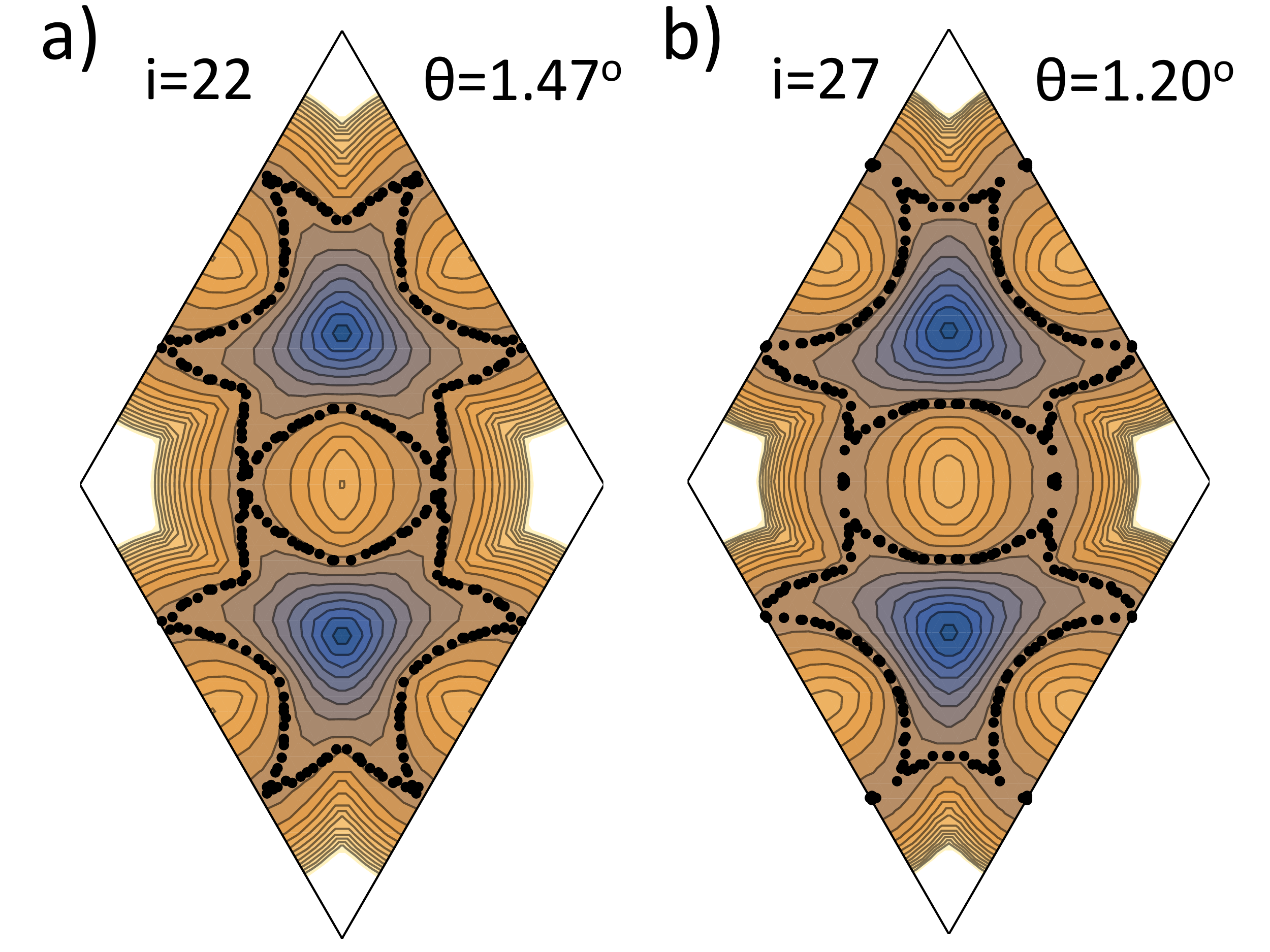}
\includegraphics[width=0.49\columnwidth]{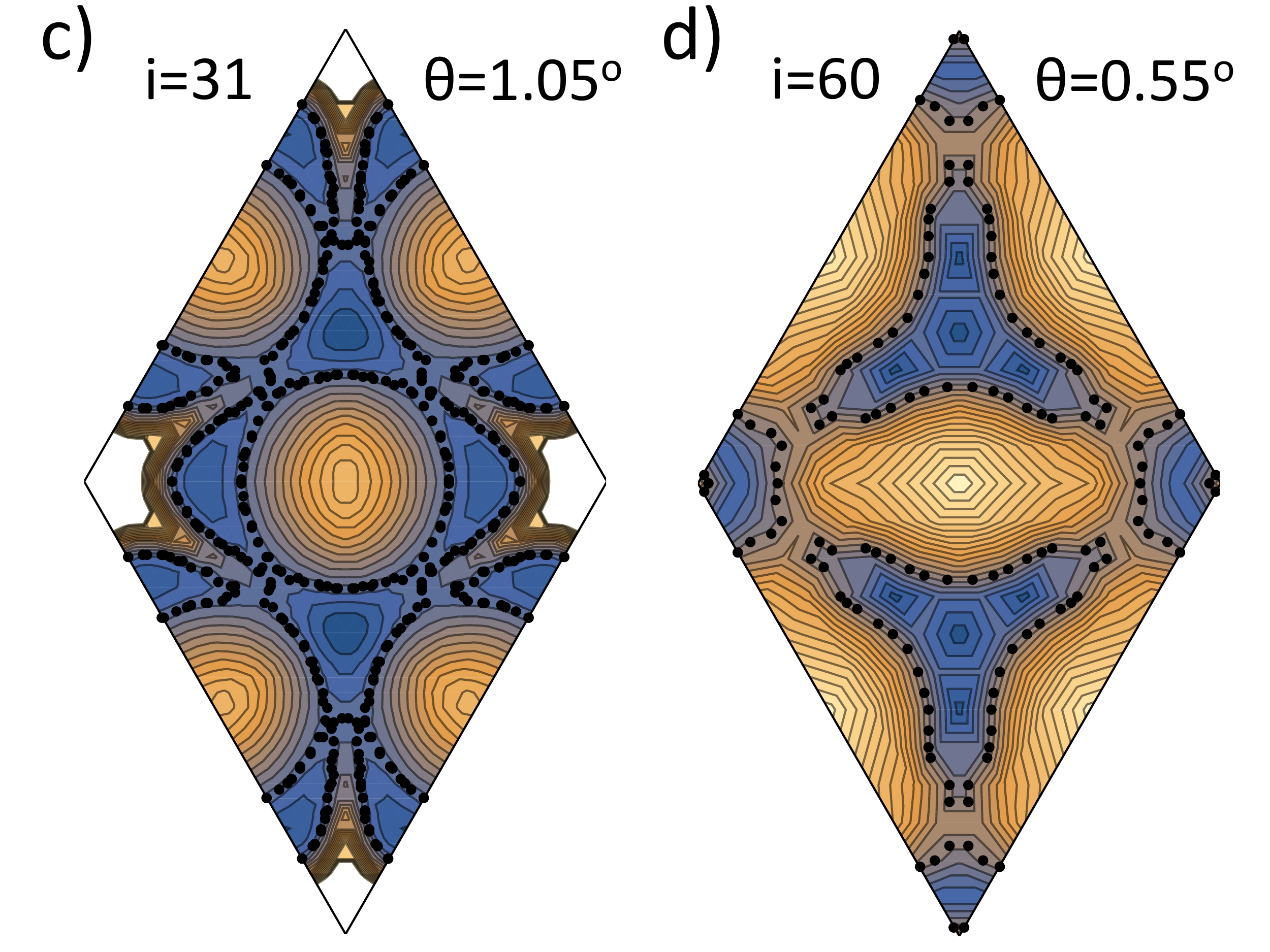}
\caption{The density plot of the highest valence bands of the relaxed lattice and the energy contour at the vHS marked by black dots. a) and b) For the commensurate twist angles with $i=22$ and $i=27$, the vHS are located on the zone boundary of the MBZ. c) and d) For the commensurate twist angle with $i=31$ and $i=60$, the vHS are located inside the MBZ, demonstrating the doubling of the vHS.
\label{vanHoveUniversality}}
\end{figure*}

\section{Filling factor at the van Hove singularity}
Superconductivity occurs around the filling factor $n=n_s/2$ where $n_s=4/A_i$ is the density needed to completely fill one band of the MBZ, with $A_i=(3i^3+3i+1)A_c$ and $A_c=a^2\sqrt{3}/2$ the area of the unit cell of single layer graphene with $a=0.246$nm. In order to efficiently apply the Kohn-Luttinger mechanism, this filling factor should be linked to the Fermi level at the vHS. In the following, we will argue that for a non-interacting band structure, this is the case for twist angles close to the magic angle. We will then present experimental evidence that suggests a renormalization of the van Hove energy in the interacting theory. 

\subsection{General remarks}
Let us start with general considerations on the filling factor and the Fermi surface at the vHS. For single layer graphene (SLG), within a simple tight-binding model including only nearest-neighbor hopping, the Fermi surface at the vHS $E_{vH}$ has a triangular form and corresponds to $n=\pm n_g/4$ where $n_g=4/A_c$ is the electron density needed to completely fill one band of the full Brillouin zone, see Fig. \ref{BZvanHove} a). 

This is modified in the case of TBG as discussed in the following. For large twist angles, the Fermi surface at the van Hove energy $E_{vH}$ can be approximated by a circle due to the isotropic Dirac cone physics expected for the highest valence band, i.e., $n\approx\pm \sqrt{3}/\pi n_s\to0.55 n_s$. For small twist angles, we assume a doubling of the vHS and thus an inversion of the triangle characterizing the Fermi surface at $E_{vH}$ of SLG. Constructing a simple geometrical model consistent with this assumption, the filling factor becomes exactly 1/2 for the Fermi energy at $E_{vH}$, i.e., $n=\pm n_s/2$, see Fig. \ref{BZvanHove} b). 

Let us now consider the density plot on the MBZ for the eigenenergies of the highest valence band within the continuous model of Ref. \cite{Lopes07}. For $\theta_{i=5}=6^\circ$, we approximately recover the simplified model described above with a circular Fermi surface even at $E_{vH}$, see Figs. \ref{BZvanHoveValley}, \ref{BZvanHovePlus}, \ref{BZvanHoveMinus} a). More interestingly, also the model for small twist angle can be approximately realised close to the magic angle with $\theta_{i=30}=1.1^\circ$, see Figs. \ref{BZvanHoveValley}, \ref{BZvanHovePlus}, \ref{BZvanHoveMinus} d).

\begin{figure}
\includegraphics[width=0.49\columnwidth]{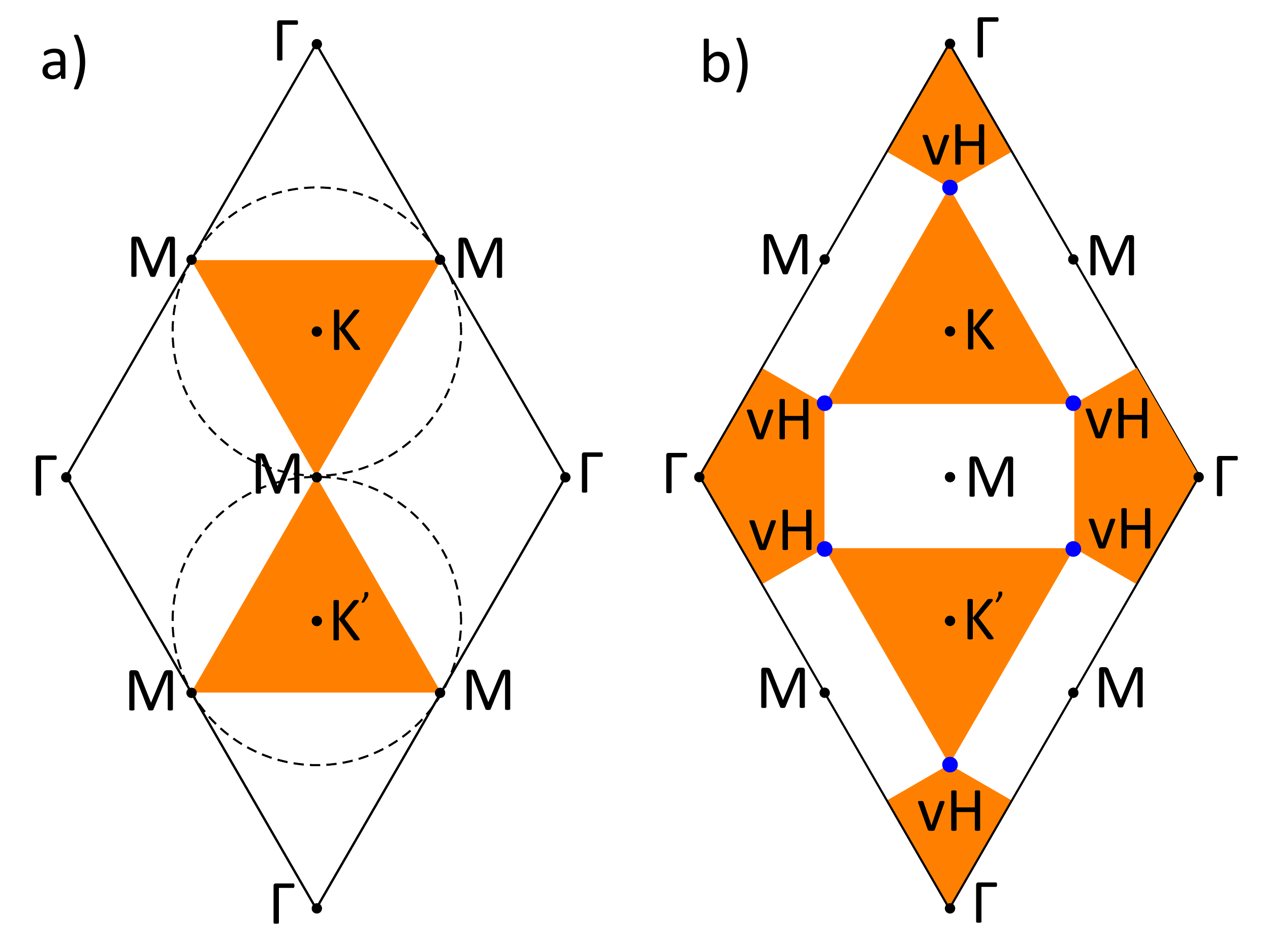}
\includegraphics[width=0.49\columnwidth]{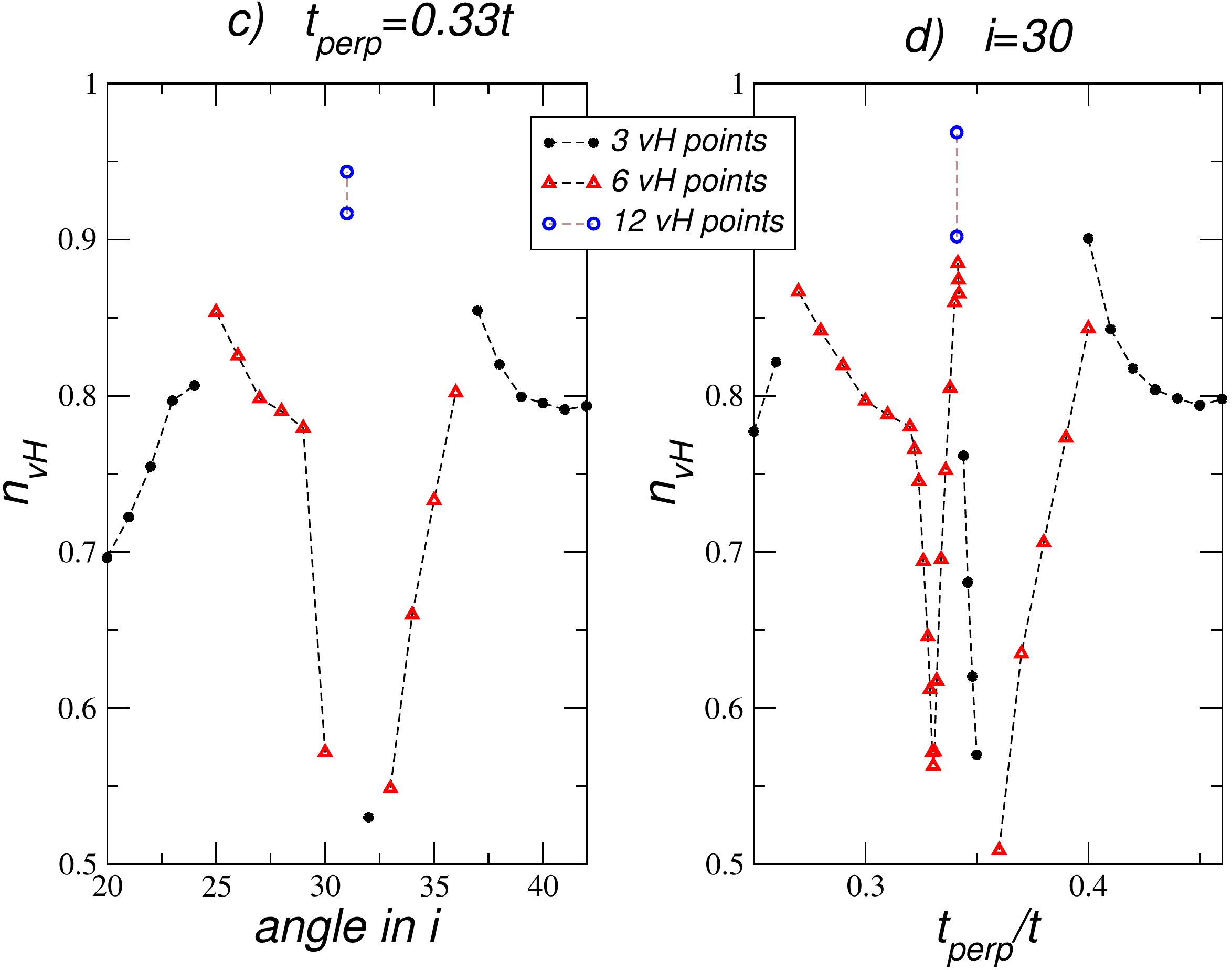}
\caption{a) Brillouin zone (BZ) and empty states (orange) for single layer graphene (SLG) with Fermi energy at the van Hove singularity of the highest valence band. Also shown the approximate Fermi level of TBG with large twist angle as dotted circles that favorably compares with Figs. \ref{BZvanHoveValley}, \ref{BZvanHovePlus}, \ref{BZvanHoveMinus} a). Black points indicate the high-symmetry points of the BZ and in both cases the van Hove singularities coincide with the $M$-points. b) Moir\'e Brillouin zone (MBZ) and empty states (orange) for TBG at a small angle with Fermi energy at the van Hove singularity. Black points again indicate the high-symmetry points of the MBZ, but now the vHS (blue dots) lie in the center of the line connecting the $\Gamma$- and the $K$-points. This Fermi surface favorably compares with Figs. \ref{BZvanHoveValley}, \ref{BZvanHovePlus}, \ref{BZvanHoveMinus} d). c) and d): Filling factor of the highest valence band at the van Hove singularity $n_{vH}$ close to the first magic angle as function of c) the twist angle in units of $i$ for fixed interlayer coupling $t_\perp=0.33t$. d) the interlayer coupling $t_\perp$ in units of the interlayer coupling $t$ for fixed twist angle $i=30$. We also indicate the number of van Hove singularities of one valley. The magic angle is given by the appearance of 12 vHS at almost degenerate energy. Bifurcations from 3 to 6 vHS at a given hopping parameter $t_\perp$ also occur at approximately the same energy.
\label{BZvanHove}}
\end{figure}

\subsection{Evolution of saddle points}
As already discussed, there occurs a doubling of the vHS at some critical angle $\theta_c^+ \approx 1.3^\circ$. But this is not the only ^^ ^^ critical" angle where the topography of the vHS changes. In Fig. \ref{BZvanHove} c), the electron density at the energy of the vHS of the highest valence band as well as the number of vHS is shown for various commensurate twist angles $i$. The same qualitative behavior (albeit more continuous) is shown in Fig. \ref{BZvanHove} d) for fixed twist angle $i=30$ and variable hopping parameter $t_\perp$. For the above analysis, the continuous model of Sec. \ref{ContinuousHamiltonian} was used and only one valley was considered. 

Let us now briefly discuss the band structure for twist angles at the magic angle and beyond. In Fig. \ref{BeyondMagic} a) and b), we show the density plots of the highest valence band for the magic angle $i=31$ and for $i=32$, as obtained from the continuous model and including both valleys, this time. Comparing these plots with the density plot at $i=30$ shown in Fig. \ref{BZvanHovePlus} d), one sees abrupt change in the electron density, i.e., the surface enclosed by the Fermi contour, and number of saddle-points as depicted in Figs. \ref{BZvanHove} c) and d). Notice also that the six lobes around the $\Gamma$-point for $i=30$ develop into six pockets around the $\Gamma$-point for $i\geq31$

In Fig. \ref{BeyondMagic} c), d), e), and f), the contours are shown where the degeneracy of the highest valence band of the two valleys occurs. For $i\leq29$, the only lines are the ones that connect the $\Gamma$ and the $K$-point. For $i\geq 30$, also a circle around the $\Gamma$-point emerges and for $i=32$, new features around the $K$-point develop. The degeneracy of the highest valence band of the two valleys is thus another indicator for the onset of the ^^ ^^ small" or ^^ ^^ magic" angle regime.

\subsection{Renormalization of the saddle points}

The above considerations apply to the non-interacting theory, but renormalization effects at the vHS can be expected in the interacting electron system. In this respect, a renormalization of the chemical potential near a vHS has been predicted and discussed in Ref. \cite{Gonzalez97}. From the experimental point of view, a strong renormalization of the level of the vHS from $\sim 2.7$ eV to $\sim 1.8$ eV has been observed in the conduction band of graphene\cite{McChesney10}.

It is therefore advisable to have some experimental input in order to establish the filling level at the vHS.
In Refs. \cite{Kim16b,Cao16}, the Hall conductivity has been measured for TBG in the small angle regime. Interestingly, an abrupt sign change has been observed at the filling factor $n=n_s/2$. A sign change in the Hall conductivity is usually associated with a change in the carrier type, i.e., from $n$ to $p$-type, or ---more generally--- with the change of the sign of the effective mass. The sign of the effective mass changes across a vHS, so that those experiments provide a strong suggestion that the vHS may be pinned to half-filling in the highest valence band of TBG. 

Let us finally note that a renormalisation of the chemical potential of the van Hove singularity has been predicted and discussed in Ref. \cite{Gonzalez97}. Also, it was also proposed that in the case of single-layer graphene (SLG), anisotropic screening around the vHS\cite{Stauber10} can lead to superconductivity via the Kohn-Luttinger (KL) mechanism\cite{Kohn65,Gonzalez08} and a strong renormalisation of the vHS from $\sim2.7$eV to $\sim1.8$eV was observed.\cite{McChesney10}

\begin{figure}
\includegraphics[width=0.49\columnwidth]{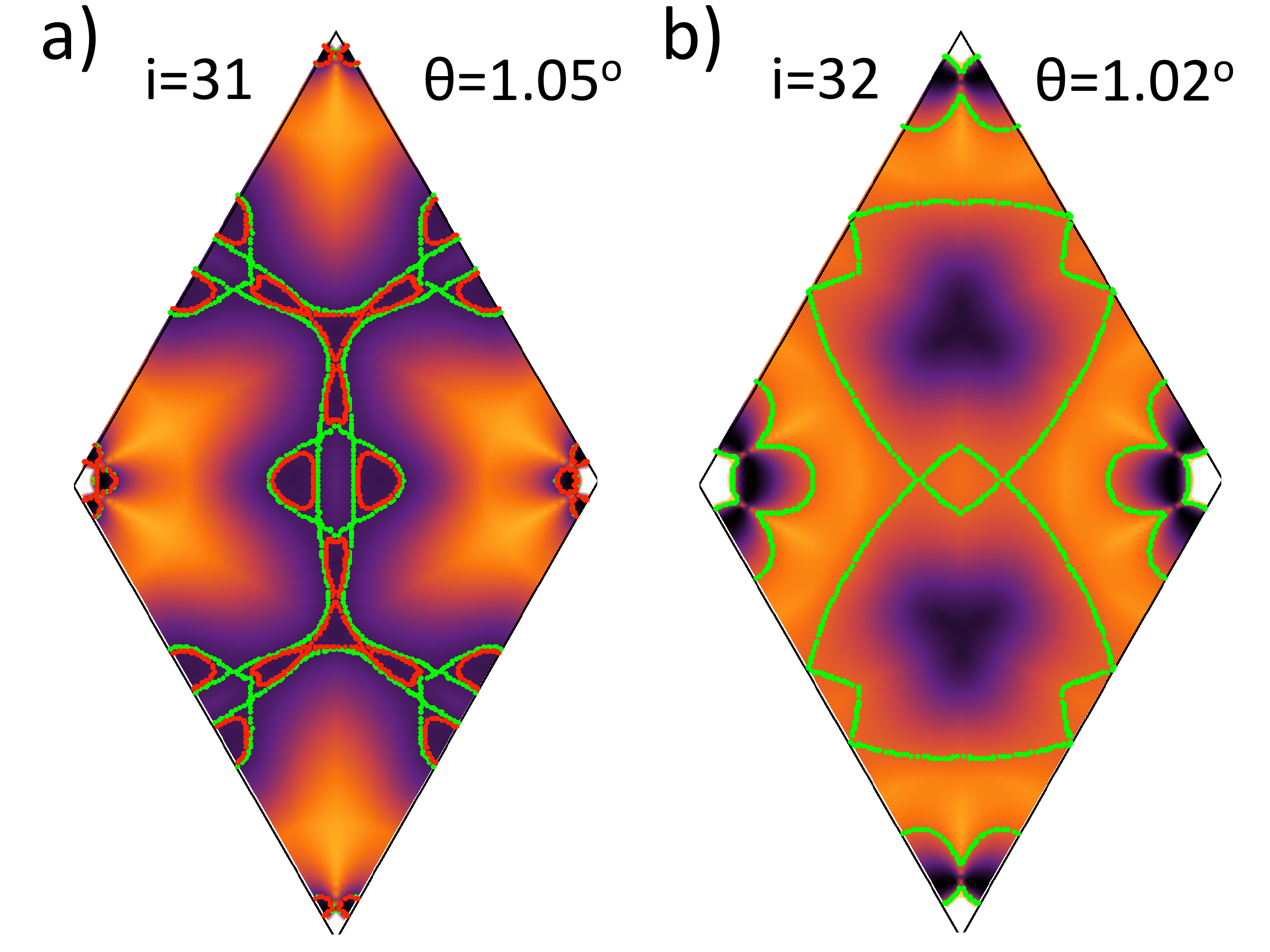}
\includegraphics[width=0.49\columnwidth]{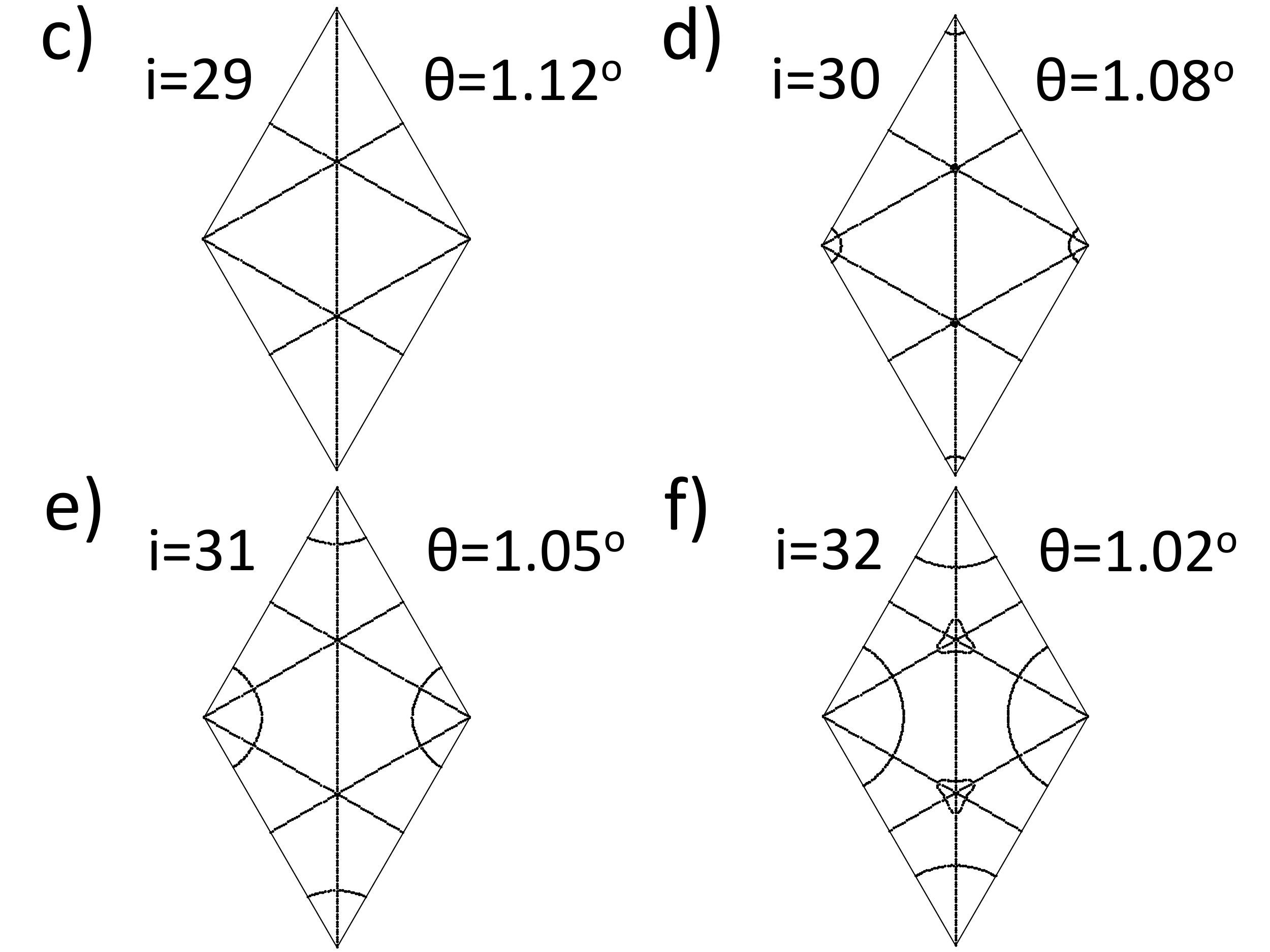}
\caption{a) and b): Density plot of the lowest valence band including both valleys for $i=31$ and $i=32$. Also shown are the contours of constant energies at the vHS. The green curve for $i=32$ corresponds to the energy at the van Hove singularity of the second highest valence band. c), d), e), and f): The degeneracy contours of the highest valence band of the two valleys. For $i\leq29$, the only degeneracies occur along the lines in the direction $\Gamma K$. For $i\geq30$, the degeneracy occurs also along a circle around the $\Gamma$-point. Finally, for $i=32$, now degeneracies emerge around the $K$-points.
\label{BeyondMagic}}
\end{figure}

 \section{Scaling near 2D van Hove singularities}

\subsection{Renormalization group analysis}

The divergent density of states at a vHS severely constrains the form of the interactions which are not irrelevant for electrons near the saddle point dispersion. This can be shown by resorting to a renormalization group (RG) method, which allows to identify the effective action of the interacting electron system at low energies. Following the steps of a Wilsonian approach, one first has to set a high-energy cutoff $\Lambda $ in the many-body theory, dictated by the microscopic length scale of the system, with the idea of integrating subsequently the modes with energy $\varepsilon $ in a thin shell $\Lambda/s < |\varepsilon| < \Lambda $. By repeated application of this procedure, one may inspect the behavior of the different interactions when the cutoff is progressively lowered, which corresponds to taking the low-energy limit of the electron system. If some of the interactions grow large in this limit, it means that the original formulation of the theory is not capturing the relevant degrees of freedom. Otherwise, some of the interactions may fade away as the electron modes are integrated out, signaling their irrelevance in the low-energy limit. When the recursive integration ends up however with a reduced number of finite interactions, one obtains in this way an effective theory with which to compute reliably response functions and other observables in the low-energy limit.

The case of an electron system with saddle-point dispersion in two dimensions adapts well to the RG approach, as the noninteracting theory already shows a scale invariant behavior under the progressive reduction of the high-energy cutoff $\Lambda $. The action of the noninteracting model can be written in terms of creation (annihilation) operators $\Psi^+_{\sigma }({\bf p})$ ($\Psi_{\sigma }({\bf p})$) for electrons with momentum ${\bf p}$ and spin $\sigma = \uparrow, \downarrow $ as 
\begin{equation}
S_0   =   \int dt d^2 p \left( i \Psi^+_{\sigma }({\bf p}) \partial_t \Psi_{\sigma }({\bf p}) 
  - \varepsilon ({\bf p})  \;  \Psi^+_{\sigma }({\bf p}) \Psi_{\sigma }({\bf p})  \right)   
\label{act0}
\end{equation}
where the dispersion is given by 
\begin{equation}
\varepsilon ({\bf p})  \approx \alpha p_x^2 - \beta p_y^2
\end{equation}
In Eq. (\ref{act0}), one assumes that the momentum is restricted within the region constrained by the high-energy cutoff $\Lambda $. After integrating out the modes in the shell with $\Lambda/s < |\varepsilon| < \Lambda $, one is just left with the modes in the range $|\varepsilon| < \Lambda/s $. In order to make the comparison with the original action, one has to rescale the new cutoff to $\Lambda $, which is made by changing variables to ${\bf p}' = s^{1/2} {\bf p}$. Indeed, the action can be restablished to its original form by means of the transformation rule
\begin{eqnarray}
\partial_{t'} & = &  s \partial_t   \label{s1}   \\
{\bf p}' & = &  s^{1/2} {\bf p}           \\
\Psi_{\sigma }' ({\bf p}) & = &  s^{-1/2} \Psi_{\sigma } ({\bf p})
\label{s3}
\end{eqnarray}   
That is, the action (\ref{act0}) for the saddle-point dispersion is a fixed-point of the RG transformations.

The important point is to check the behavior of the $e$-$e$ interactions under the scale transformation (\ref{s1})-(\ref{s3}). We can write the part of the action containing a most general type of four-fermion interaction as 
\begin{equation}
S_{\rm int} = \int dt d^2 p_1 d^2 p_2 d^2 p_3 d^2 p_4  \; U ({\bf p}_1, {\bf p}_2, {\bf p}_3, {\bf p}_4) \; 
  \Psi^+_{\sigma }({\bf p}_1)  \Psi^+_{\sigma'}({\bf p}_2) 
  \Psi_{\sigma'} ({\bf p}_4) \Psi_{\sigma } ({\bf p}_3) \; \delta ( {\bf p}_1 + {\bf p}_2 - {\bf p}_3 - {\bf p}_4 ) 
\label{act}
\end{equation}
where $U ({\bf p}_1, {\bf p}_2, {\bf p}_3, {\bf p}_4)$ stands for the interaction potential. After integration of the modes at the high-energy cutoff, one has to make again the change of variables to redefine the cutoff from $\Lambda/s $ to $\Lambda $. Then, one observes that the action (\ref{act}) can remain invariant under the transformation (\ref{s1})-(\ref{s3}), provided that the interaction potential is kept constant. In general, the potential $U$ may be expanded in powers of the momenta, and it is clear that only the zeroth-order term gives rise to an interaction leaving (\ref{act}) scale invariant. Higher-order powers of ${\bf p}$ in the expansion are going to be modified by negative powers of $s$ after the change of variables (\ref{s1})-(\ref{s3}), meaning that they become irrelevant in the low-energy limit $s \rightarrow \infty $. 

The conclusion is that, in the electron system with saddle-point dispersion, only the constant (zeroth-order) term of the interaction potential in momentum space can give rise to sensible effects in the low-energy regime of the interacting theory. Correspondingly, this amounts to say that just an effective local interaction in real space is needed to describe the low-energy physics of the saddle-point dispersion, which can be interpreted as the influence that the divergent density of states has to limit drastically the range of the effective interaction.

\subsection{Electronic instabilities}

While the above analysis has been carried out at the level of the action in (\ref{act0}) and (\ref{act}), the RG method becomes also quite powerful to determine the relevance or irrelevance of the different quantum corrections in the many-body theory. The approach follows the same procedure discussed in the above section, with the aim of identifying the terms which may survive in the full effective action of the theory as $s \rightarrow \infty $. In this low-energy limit, it may turn out that some of the effective interactions grow large, which has to be understood as the signal of a low-energy instability in the system. In this regard, the RG approach is one of the most reliable methods to study the competition between different instabilities in a low-dimensional electron system, as it considers different quantum corrections on equal footing when applying the recursive integration of high-energy degrees of freedom.     

A well-known example of electronic instability described in the framework of the RG approach is the case of BCS superconductivity. The origin of that electronic instability lies in the divergence of the corrections to electron scattering in the so-called BCS channel, when the momenta of the two incoming electrons add to zero as represented in Fig. \ref{vert}(a). The lowest-order correction to the four-fermion vertex corresponds to the diagram shown in Fig. \ref{loop}(a), which is built with a particle-particle susceptibility diverging in general like the density of states times $\log (\Lambda )$. If we denote the vertex with BCS kinematics by $V$ (without paying attention at this point to possible dependence on the momenta), we can write the variation under a reduction of the cutoff from $\Lambda $ to $\Lambda/s $ as
\begin{equation}
d V = c \: n(\Lambda ) \: \frac{d \Lambda }{\Lambda } \: V^2
\label{diff}
\end{equation}
where $c$ is a constant and $n(\Lambda )$ stands for the density of states. From (\ref{diff}) we may obtain the evolution of the BCS vertex under the progressive reduction of the cutoff, which is encoded in the RG equation
\begin{equation}
\Lambda  \frac{\partial V}{\partial \Lambda } = c \: n(\Lambda ) \: V^2
\label{flow}
\end{equation}

\begin{figure*}[h]
\includegraphics[width=0.3\columnwidth]{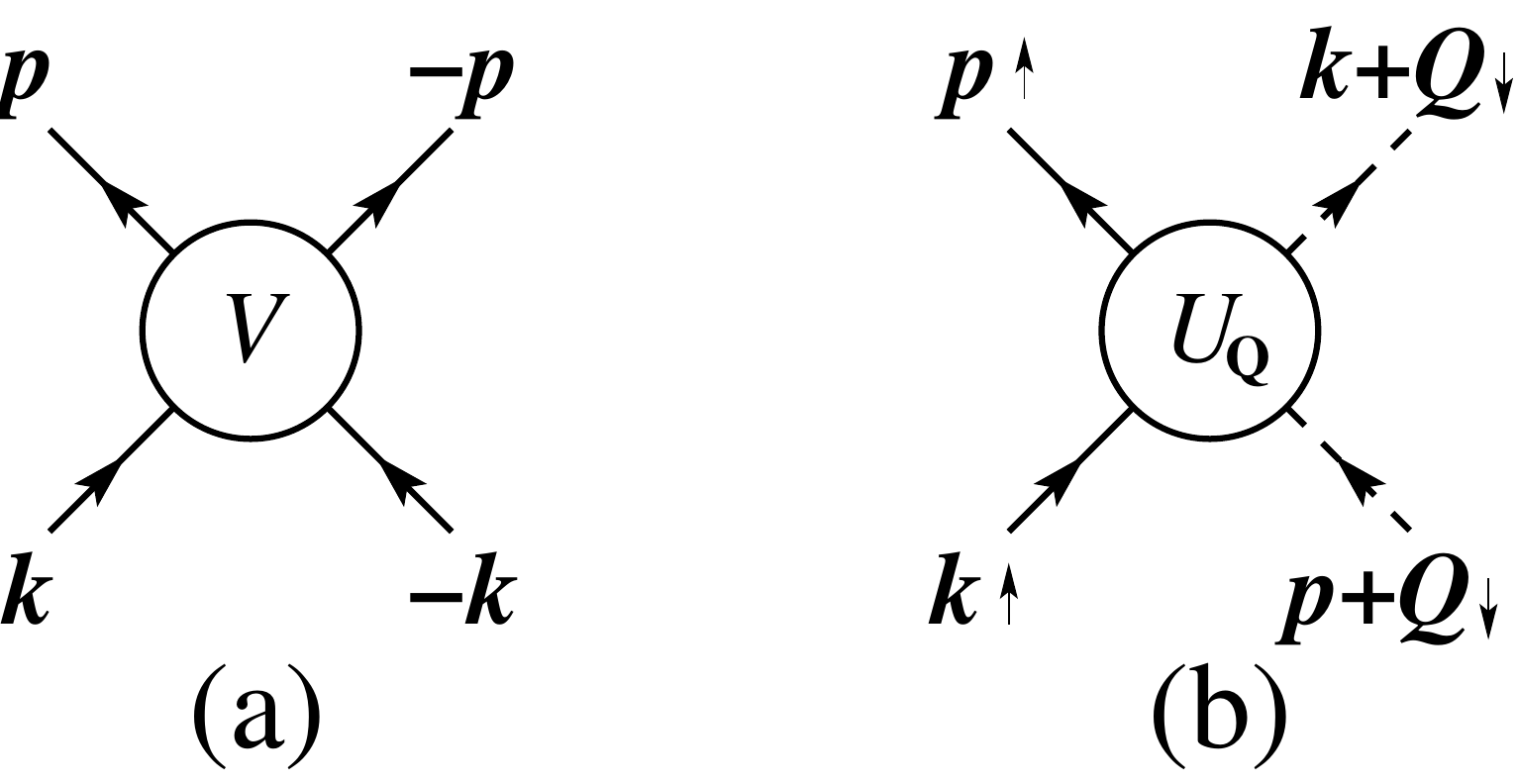}
\caption{Four-fermion vertices with the kinematics relevant for the scattering of electrons near a saddle-point dispersion. 
\label{vert}}
\end{figure*}

The resolution of (\ref{flow}) leads to a steady decrease of the vertex in the limit $\Lambda \rightarrow 0$ when the original value of $V$ at the high-energy cutoff is set by a bare repulsive interaction, in such a way that $V(\Lambda ) > 0$. If however the dominant contribution comes from an attractive interaction at the cutoff $\Lambda_0$, so that $V(\Lambda_0) < 0$, the solution of (\ref{flow}) leads to a low-energy pairing instability in the system, with the onset marked by the singularity of the RG flow at the scale
\begin{equation}
\omega_c \approx \Lambda_0 \exp \left(-\frac{1}{c \: n |V(\Lambda_0)|} \right)
\end{equation}
(assuming a constant density of states $n$). In conventional electron systems, the dominant attractive interaction may come from the coupling to phonons, for energies below the frequency of the phonon branch. In this regard, the Kohn-Luttinger mechanism represents an alternative which relies on the possibility to induce an effective attraction from the original Coulomb repulsion. This is the mechanism discussed in detail in the main text, where we have paid attention to the precise dependence of the vertex $V$ on the momenta of the incoming and outgoing electrons. When carrying out the analysis in terms of the different harmonics pertinent to the symmetry of the Fermi line, we have seen that an effective attraction indeed develops in very specific channels, as a consequence of the highly anisotropic screening of the Coulomb interaction in our model.

\begin{figure*}[h]
\includegraphics[width=0.4\columnwidth]{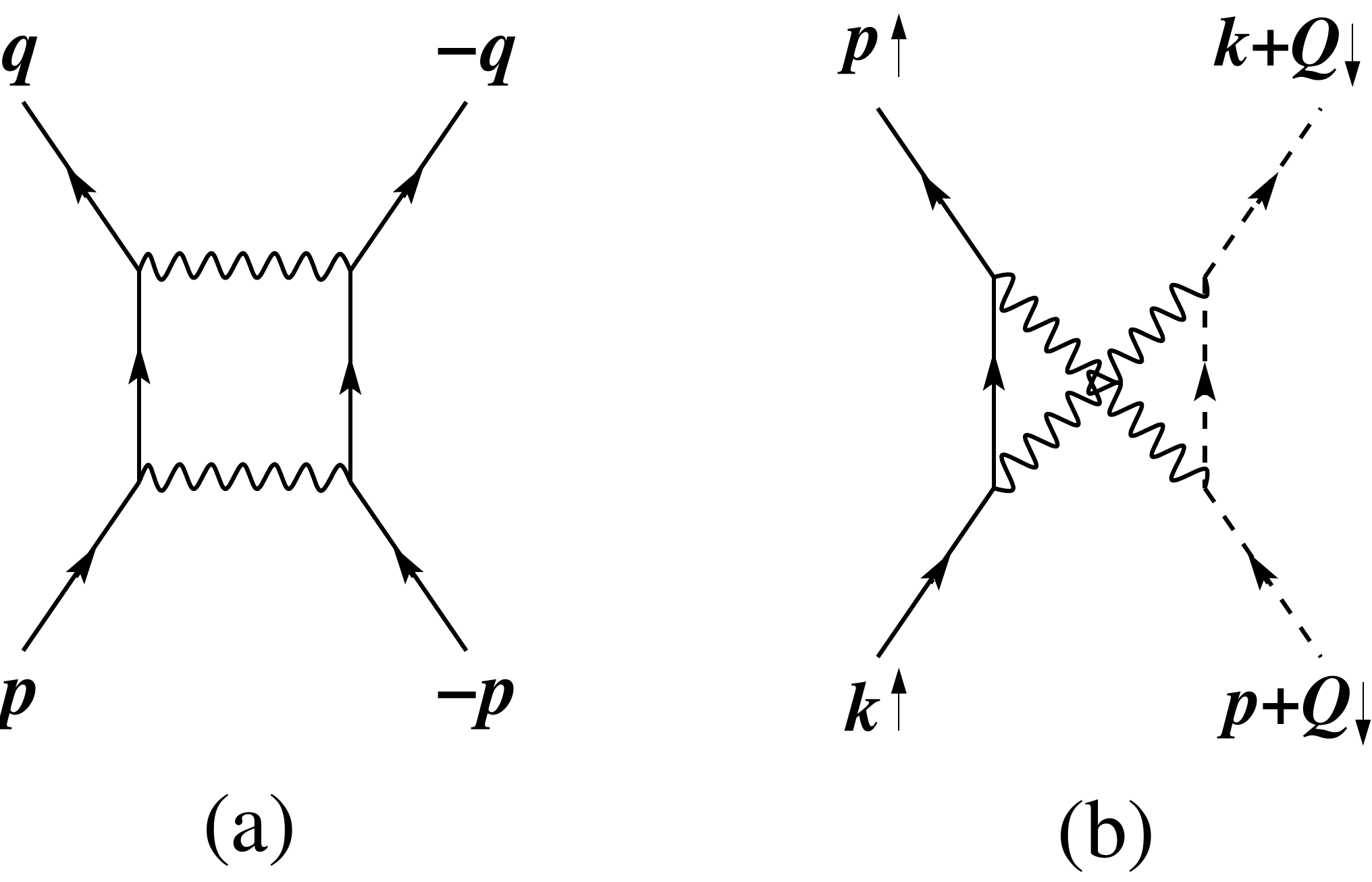}
\caption{Lowest-order corrections to the vertices $V$ and $U_{\bf Q}$. The wavy lines stand in (b) for the interaction $U_{\bf Q}$ between electron currents with opposite spin and located at different spots separated by momentum ${\bf Q}$. 
\label{loop}}
\end{figure*}

At this point, one may look for other type of vertices which can grow large at low energies. This requires the existence of some divergent susceptibility apart from the particle-particle susceptibility. In the model with saddle-point dispersion, the divergent density of states gives rise indeed to particle-hole susceptibilities that diverge as $\log (\Lambda )$ in the limit $\Lambda \rightarrow 0$. This kind of behavior is even enhanced when there is nesting of the Fermi line, that is, parallel segments which are connected by a constant translation vector ${\bf Q}$ that exchanges the particle and hole character of the electronic states. This is actually what happens in the Fermi line of the twisted bilayers we have considered, as can be seen in Fig. 2(b) of the main text. The particle-hole susceptibility $\chi_{\rm ph}$ for momentum transfer equal to ${\bf Q}$ has the behavior
\begin{equation}
\chi_{\rm ph} ({\bf Q},\omega) \approx  c' \log (\Lambda /\omega )
\label{ph}
\end{equation}
with a constant $c'$ that gives a measure of the enhancement due to the nesting of the Fermi line. This prefactor is precisely captured in the calculations reported in the main text, having for instance a reflection in the peak of the susceptibility represented as a function of the momentum in the inset of Fig. 2(a).

The study of the low-energies instabilities in the case of nesting around vHSs can be found in Refs. \onlinecite{Alvarez98,Gonzalez02,Gonzalez03}, so we give here a brief account leading to the main results. It can be shown that the dominant instability in the particle-hole channel points at the development of spin-density-wave order, which can be analyzed from the response function of the spin operator
\begin{equation}
 S_j ({\bf Q}) = \sum_{k} 
\Psi^{+}_{\sigma} ({\bf k} + {\bf Q})
       \sigma^{\sigma \sigma '}_j
     \Psi_{\sigma '} ({\bf k}) \;\;\;\;\;\;\; j = x,y,z
\label{spinq}
\end{equation}
where $\sigma_j$ stand for the Pauli matrices. The simplest way to evaluate the instability in the spin sector is to look at the response function $R_x ({\bf Q},\omega)$ for $S_x ({\bf Q})$ (the other spin response functions having the same behavior due to rotational invariance). That is given to lowest order by the diagram in Fig. \ref{rx}, where the four-fermion interaction corresponds to the vertex with the kinematics shown in Fig. \ref{vert}(b). We observe that this four-fermion vertex, that we denote by $U_{\bf Q}$, is corrected by the particle-hole susceptibility (\ref{ph}) as shown in the diagram of Fig. \ref{loop}(b). As in the case of the BCS vertex, $U_{\bf Q}$ develops then a logarithmic dependence on the cutoff $\Lambda $, which is encoded (according to the diagram of the figure) in the RG equation
\begin{equation}
\Lambda  \frac{\partial U_{\bf Q}}{\partial \Lambda } = -c' \: U_{\bf Q}^2
\label{qflow}
\end{equation}
The important point is that the solution of Eq. \ref{qflow} leads to a low-energy instability, marked by the singular behavior of the vertex at a scale
\begin{equation}
\omega_c' \approx \Lambda_0 \exp \left(-\frac{1}{c' \: U_{\bf Q}(\Lambda_0)} \right)
\end{equation}

\begin{figure*}[h]
\includegraphics[width=0.2\columnwidth]{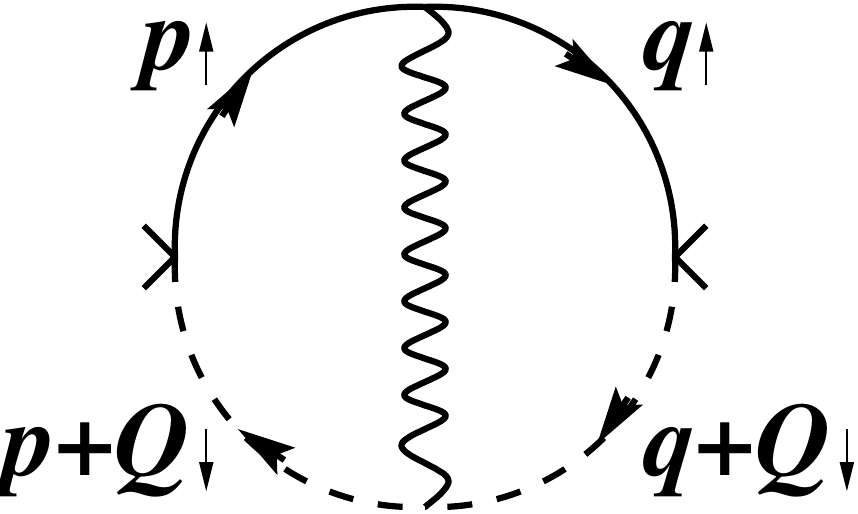}
\caption{Lowest-order correction to the spin response function $R_x ({\bf Q},\omega)$. The wavy line stands for the interaction $U_{\bf Q}$ between electron currents with opposite spin and located at different spots separated by momentum ${\bf Q}$. 
\label{rx}}
\end{figure*}

We recall that the spin response function $R_x ({\bf Q},\omega)$ obeys its own RG equation\cite{Gonzalez02,Gonzalez03}, which encodes the iteration of the interaction $U_{\bf Q}$ in the diagram of Fig. \ref{rx} to read
\begin{equation}
\frac{\partial R_{x}}{\partial \Lambda} = - 2 c'' \frac{1}{\Lambda}  
        - c'' \frac{1}{\Lambda}  U_{\bf Q} R_{x}
\label{x}
\end{equation}
The resolution of (\ref{qflow}) and (\ref{x}) goes well beyond perturbation theory, as it amounts to a partial sum of the perturbative expansion. The result is that the spin response function inherits the singularity at $\omega_c' $, signaling the development of a nonvanishing expectation value of the spin operator, $\langle S_i ({\bf Q}) \rangle \neq 0$. The vector ${\bf Q}$ of the spin-density wave is dictated by the momentum at which the particle-hole susceptibility is enhanced as a consequence of the nesting of the Fermi line. For the particular twisted bilayers we have considered, such a momentum can be easily identified by the peak in $\chi_{\rm ph}$ represented as a function of the momentum, as it is shown for instance in the inset of Fig. 2(a) in the main text.

After all, the resolution of the competition between the superconducting and the spin-density-wave instability amounts to a comparison between the solution of Eqs. (\ref{flow}) and (\ref{qflow}), in order to see which vertex function develops first a singularity as $\Lambda \rightarrow 0$. The results of this analysis are reported in the main text, and they are summarized in the phase diagram of Fig. 3(a). The main conclusion is that the superconducting instability is able to prevail in the regime connected to weak-coupling, due to the large enhancement of the instability from the divergent density of states, while large values of the bare repulsion tend to favor instead the spin-density-wave instability before the pairing instability has time to develop.

\end{widetext}
 
\bibliography{KohnLuttinger.bib}
\end{document}